%% file: main.tex
\documentclass[acmsmall]{acmart}

\AtBeginDocument{%
  }

\setcopyright{acmlicensed}
\acmJournal{PACMMOD}
\acmYear{2025} \acmVolume{3} \acmNumber{6 (SIGMOD)} \acmArticle{356} \acmMonth{12}\acmDOI{10.1145/3769821}
\acmISBN{978-1-4503-XXXX-X/2018/06}

\input{header.tex}
\begin{document}

\title{Reliable and Private Utility Signaling for Data Markets}



\author{Li Peng}
\affiliation{
  \institution{State Key Laboratory of Blockchain and Data Security, Zhejiang University; Alibaba Group}
  \country{China}
}
\email{jerry.pl@alibaba-inc.com}

\author{Jiayao Zhang}
\affiliation{
  \institution{State Key Laboratory of Blockchain and Data Security, Zhejiang University}
  \country{China}
}
\email{jiayaozhang@zju.edu.cn}

\author{Yihang Wu}
\affiliation{
  \institution{State Key Laboratory of Blockchain and Data Security, Zhejiang University}
  \country{China}
}
\email{yhwu_is@zju.edu.cn}

\author{Weiran Liu}
\affiliation{
  \institution{Alibaba Group}
  \country{China}
}
\email{weiran.lwr@alibaba-inc.com}

\author{Jinfei Liu}
\authornote{Corresponding author.}
\affiliation{
  \institution{State Key Laboratory of Blockchain and Data Security, Zhejiang University; Hangzhou High-Tech Zone (Binjiang) Blockchain and Data Security Research Institute}
  \country{China}
}
\email{jinfeiliu@zju.edu.cn}

\author{Zheng Yan}
\affiliation{
  \institution{Xidian University}
  \country{China}
}
\email{zyan@xidian.edu.cn}

\author{Kui Ren}
\affiliation{
  \institution{State Key Laboratory of Blockchain and Data Security, Zhejiang University}
  \country{China}
}
\email{kuiren@zju.edu.cn}

\author{Lei Zhang}
\affiliation{
  \institution{Alibaba Group}
  \country{China}
}
\email{zongchao.zl@taobao.com}

\author{Lin Qu}
\affiliation{
  \institution{Alibaba Group}
  \country{China}
}
\email{xide.ql@taobao.com}

\renewcommand{\shortauthors}{Li Peng et al.}

\begin{abstract}
The explosive growth of data has highlighted its critical role in driving economic growth through data marketplaces, which enable extensive data sharing and access to high-quality datasets. To support effective trading, signaling mechanisms provide participants with information about data products before transactions, enabling informed decisions and facilitating trading. However, due to the inherent free-duplication nature of data, commonly practiced signaling methods face a dilemma between privacy and reliability, undermining the effectiveness of signals in guiding decision-making.

To address this, this paper explores the benefits and develops a non-TCP-based construction for a \textit{desirable} signaling mechanism that simultaneously ensures privacy and reliability. We begin by formally defining the desirable utility signaling mechanism and proving its ability to prevent suboptimal decisions for both participants and facilitate informed data trading. To design a protocol to realize its functionality, we propose leveraging maliciously secure multi-party computation (MPC) to ensure the privacy and robustness of signal computation and introduce an MPC-based hash verification scheme to ensure input reliability. In multi-seller scenarios requiring fair data valuation, we further explore the design and optimization of the MPC-based KNN-Shapley method with improved efficiency. Rigorous experiments demonstrate the efficiency and practicality of our approach. 

\end{abstract}


\received{April 2025}
\received[revised]{July 2025}
\received[accepted]{August 2025}

\maketitle

\vspace{-0.1in}
\section{Introduction}
\label{sec:intro}

The last decades have witnessed significant value and explosive growth of data. Data marketplaces create opportunities for accessing to diverse and high-quality data, making data trading more convenient~\cite{pei2020survey,zhang2024survey}. However, reaching a data transaction is challenging when data sellers or buyers lack pre-contractual knowledge of the actual value of the data. Sellers may "sell at a loss" due to underestimation of the data value, while buyers may suffer from "buyer’s remorse" due to overestimation. This informational absence undermines the incentives for data sellers or buyers to engage in transactions.

To bridge the gap, signaling \cite{connelly2011signaling}, a widely studied method in economics, is employed to facilitate effective data trading in data markets \cite{zhang2024survey}. In the context of data transactions, a signal refers to the utility of a dataset, defined as the performance of the dataset on a buyer-specified task and metric. The signal indicates the value of the selling data, helping parties to make an informed decision, as illustrated in the following example.

\setcounter{example}{0}
\begin{example}
    (Precision Marketing) Consider a scenario in which Instagram, the data seller, maintains a dataset $\mathcal{D} := (\textsf{idA})$ consisting of users interested in photography. Amazon, the data buyer, aims to identify potential camera buyers among its own user base $\mathcal{T} := (\textsf{idB})$ and deliver targeted advertisements to them by leveraging Instagram's data. To assess the value of Instagram’s dataset before purchase, Amazon requests a metric $\nu$, defined as the size of the intersection between $\mathcal{D}$ and $\mathcal{T}$, \ie $\nu \gets |\mathcal{D} \cap \mathcal{T}|$. A larger $\nu$ indicates that Instagram’s data can help Amazon identify more relevant users, potentially leading to higher business value for promoting sales. Thus, the signal $\nu$ reflects the value of Instagram's data for Amazon and helps Amazon to make an informed decision.
\end{example}

Given the importance of a signal on data market, we focus on how the signal is computed and conveyed, which is called a signaling method. Commonly practiced signaling methods for data transactions can be generally categorized based on where utility evaluation occurs: \textit{cheap talk} \cite{farrell1996cheap} and \textit{costly signaling} \cite{chen2022selling}. 


\begin{example}
(Cheap talk \cite{farrell1996cheap}) A seller assesses the dataset utility using a buyer-specific metric (\eg accuracy) and reports it as a cheap signal to the buyer.
\end{example}

\begin{example}
(Costly signaling \cite{chen2022selling}) A seller shares the dataset directly with the buyer as a costly signal, allowing the buyer to assess and report the dataset utility.
\end{example}

Both cheap talk and costly signaling fall short in scenarios involving rational yet untrustful sellers. Buyers, assumed to be trustful as their primary goal is to evaluate dataset utility, rely on accurate signals to make decisions. Malicious behavior that distorts the true utility assessment may mislead the buyer, potentially leading to a loss of transaction payoff or transaction failure. In cheap talk, a rational yet distrustful seller may inflate or fabricate utility metrics to send a mistaken signal for increasing payoff, since buyers cannot verify her claim. Meanwhile, costly signaling allows buyers to directly evaluate the dataset, ensuring a reliable utility assessment while disclosing the dataset before purchase. This disclosure undermines the value of dataset being sold due to the inherent free-duplication nature of data, which is known as the Arrow information paradox \cite{arrow1972economic}. These limitations highlight the need for a desirable signal mechanism in which participants input their data and receive the desired output utility without concerns about its trustworthiness or the risk of input data leakage. That is, ensuring both \textit{reliability} and \textit{privacy} (see \S\ref{sec:problem_formulation} for formal definition).

\partitle{Challenges}
One potential solution to design a desirable signal mechanism is to introduce a trusted central party (TCP) (\eg a trusted broker) \cite{castro2023data}. The sellers and buyers transmit data and utility metrics to a TCP, respectively, and the TCP calculates and sends utility signals. However, TCP is often unavailable or impractical in many real-world scenarios. To address this, a non-TCP-based signaling mechanism is required, but several challenges exist.

\begin{itemize}
    \item \textit{Benefit analysis.} 
    In data trading scenarios, both the seller and the buyer are motivated to access the data value to facilitate informed pricing or decision-making \cite{pei2020survey}. While a desirable signaling mechanism that delivers utility is envisioned, its impact on each participant is complex.
    The seller's payoff is an indirect function of the buyer's utility, which in turn depends on the seller's pricing function and the signaled utility. This interdependence complicates analyzing the benefits of signaling. The first challenge is: \textit{How to analyze the impact of a desirable signaling mechanism on the decision-making and payoff of the buyer and seller in data trading?}

    \item \textit{Construction of a signaling protocol.} 
    The desired signaling protocol must guarantee verifiable dataset inputs, reliable utility computation, and privacy protection for both parties’ data without relying on TCPs. However, directly applying conventional cryptographic techniques, such as secure multi-party computation (MPC), falls short of simultaneously meeting these unique needs. The second challenge is: \textit{How to design a utility signaling protocol that realizes the functionality of a desirable utility signaling without relying on TCPs?}

    \item \textit{Fair utility allocation among multiple sellers.} 
    In scenarios involving multiple sellers (contributors), it is critical to equitably reflect each contributor's impact on the overall dataset utility for fair pricing or revenue allocation \cite{agarwal2019marketplace}. 
    The third challenge is: \textit{How to develop an efficient method that achieves fair utility allocation among sellers without relying on TCPs?}
\end{itemize}

\begin{figure}[t!]
    \centering
    \includegraphics[width=0.5\linewidth]{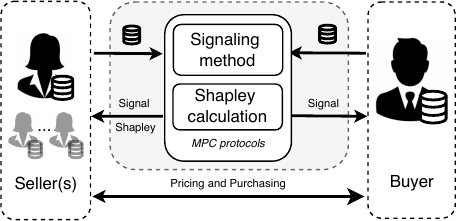}
    \vspace{-0.1in}
    \caption{System architecture.}
    \vspace{-0.2in}
    \label{fig:system_model}
\end{figure}

\partitle{Contribution} 
This paper bridges the above gaps by formulating, analyzing, and constructing the desirable utility signaling mechanism and Shapley calculation method. The system architecture is shown in Fig. \ref{fig:system_model}, which contains one data buyer and potentially multiple data sellers, and no TCP exists. Before data purchasing, they interact with each other through a signaling protocol to obtain data utility to make informed pricing or purchasing decision. A Shapley calculation method is used to fairly allocate the utility among multiple sellers. Specifically, we make the following contributions.

First, to investigate the benefits of the desirable utility signaling mechanism, we analyze a three-stage data trading process with a posted-price mechanism. By carefully modeling and introducing mild assumptions to make the payoff analysis tractable and get intuitive results, we demonstrate that a desirable utility signaling mechanism $\mathcal{M}$ benefits the data seller, the data buyer, and overall social welfare. We show that prior access to utility information effectively prevents suboptimal decisions for both participants. In addition, the true payoff of the buyer can be ensured non-negative by employing $\mathcal{M}$. This encourages successful transactions among risk-averse participants, thereby improving social welfare.

Second, to construct a protocol that implements the functionality of $\mathcal{M}$ while meeting the unique requirement of protection, we propose a novel approach based on secure multi-party computation. Trivially employing semi-honest secure MPC fails to ensure the reliability of a signal. To address this, we decouple the problem of reliability into two components: robustness of computation (RoC) and authentication of input (AoI). RoC can be achieved through the verifiability of maliciously secure MPC. For AoI, we introduce a novel hash verification scheme that leverages the collision-resistance property of hash functions to ensure input correctness. This approach provides a rigorous guarantee while incurs a linear overhead. To enhance scalability, we further propose an approximate verification approach based on sampling and Merkle hash trees, which reduces the overhead to be \textit{constant}.

Third, in scenarios involving multiple sellers, we investigate the concrete design of an MPC-based KNN-Shapley method to enable data valuation and revenue allocation, given that the Shapley value is a widely adopted method for equitable revenue distribution \cite{shapley1953value}. The focus of KNN-Shapley is motivated by its comparable accuracy and enhanced efficiency relative to the standard Shapley methods \cite{jia2019efficient}. We explore the concrete design and optimization of the MPC-based KNN-Shapley method, and exploit the use of a Radix sorting-based permutation that improves upon conventional bitonic sorting-based solutions.

The main contributions of this paper are summarized as follows.

\begin{itemize}
    \item We analyze the advantages of a desirable utility signaling in facilitating informed data trading.
    \item We construct protocols to implement the functionality of the desirable utility signaling mechanism, offering a trade-off between the rigor of guarantees and efficiency. 
    \item We introduce, for the first time, an MPC-based KNN Shapley protocol for fair allocation in a non-TCP setting, with optimization techniques applied to enhance its efficiency.
    \item Rigorous experiments based on MP-SPDZ are conducted to demonstrate the concrete efficiency of our schemes.
\end{itemize}

\partitle{Outline} The rest of the paper is organized as follows. \S\ref{sec:related_work} reviews the related work on data markets and integrity verification techniques. \S\ref{sec:preliminaries} introduces the background knowledge of realizing a desired signaling protocol, including secure multi-party computation, hash verification protocols, and Shapley value calculation. \S\ref{sec:problem_formulation_benefits} formally formulates the problem and demonstrates the benefits of a desired signaling method to motivate our solution better. 
In \S\ref{sec:construction}, we start from a basic protocol that falls short in protecting reliability, and then improve it with MPC-based hash verification methods to achieve full protection. To further support fair allocation, \S\ref{sec:private_knn_shapley} studies the design and optimization of MPC-based KNN-Shapley computation. \S\ref{sec:experiments} presents the experimental results. We conclude the paper and present future directions in \S\ref{sec:conclusion}.

\vspace{-0.05in}
\section{Related Work}
\label{sec:related_work}

\subsection{Data Markets and Signaling}
\vspace{-0.0in}

A data marketplace generally involves interactions
between data buyers and data sellers \cite{zhang2024survey,pei2020survey,castro2022protecting,sun2024profit}. 
Agarwal \etal \cite{agarwal2019marketplace} designed a data marketplace
where multiple sellers sell data and buyers come with their own ML models. Chen \etal \cite{chen2019towards} proposed a model-based pricing market where a trained ML model is sold, based on the buyer’s interest. Liu \etal \cite{liu2021dealer} proposes the first end-to-end model marketplace with differential privacy Dealer. They all assume a trusted marketplace (or broker) can access the seller's data or model for pricing based on the buyer's budget or interest. To signal participants before trading, Chen \etal \cite{chen2022selling} designs a costly signaling framework where the seller releases a portion of data to the buyers to benchmark an approximate utility as a signal. However, their goal is to study the revenue maximization of the seller, while the costly signal incurs additional privacy costs and limits seller revenue. 
Later, Fernandez \cite{castro2023data} proposes data-sharing consortia.
This work uses a trusted platform to perform tasks and signal data value to each participant, while ensuring that no data is released before transaction. Unlike these works, our work focuses on signal utility in end-to-end data trading scenarios where no TCP is present. We specifically aim to ensure signal reliability and privacy and explore its impact on both participants.

\vspace{-0in}
\subsection{Integrity Verification of Computation}
\vspace{-0in}
To ensure the reliability of computation without relying on TCP, a relevant line of research uses zero-knowledge proofs (ZKPs) \cite{goldwasser2019knowledge}. A ZKP enables a prover to generate a proof that convinces a verifier that the result of a public function $f$ on the public input $x$ and secret input $w$ of the prover is $y=f(x,w)$. By adopting the "commit-and-prove'' paradigm \cite{cramer1997linear}, ZKP-based methods can ensure the integrity of computation with respect to the private input $x$ \cite{zhang2020zero,weng2021mystique,zhang2017vsql,li2023zksql}.
However, such approaches assume that only the prover provides private input $w$, which differs from our setting where the buyer (the verifier) also has private input (detailed in \S\ref{sec:privacy_model}). ZKP systems often incur high computational costs, typically due to complex algebraic computations. An alternative line of work leverages trusted hardware enclaves \cite{schuster2015vc3,bittau2017prochlo}. These solutions process data in plaintext within an isolated environment (\eg Intel SGX), ensuring data privacy and integrity of computation. Nevertheless, these approaches require additional trust assumptions and suffer from side-channel attacks \cite{van2018foreshadow,wang2017leaky}. Fully homomorphic encryption (FHE) \cite{gentry2009fully} also aims to ensure data privacy. However, ensuring reliability against malicious behavior entails significant overhead \cite{gennaro2013fully, fiore2020boosting}.

\vspace{-0in}
\section{Preliminaries}
\label{sec:preliminaries}
\vspace{-0in}

In this section, we present the necessary technical background. The signaling process faces not only risks \textit{during} the utility computation,  but also risks \textit{outside} of utility computation, \ie risks stemming from the input data. To address these threats, we leverage MPC-based techniques and hash verification protocols. In addition, Shapley values are adopted for fair utility allocation.
Frequently used notation are summarized in Appendix \ref{appendix:notation}.

\subsection{Secure Multi-party Computation (MPC)}
\label{sec:pre_mpc}
\vspace{-0.1in}

\partitle{Security definition}
We introduce the idea of secure computation in the MPC model \cite{goldreich2001foundations}. In this model, each participating entity possesses some private data and agrees to engage in the computation $f$ jointly through a predetermined \textit{protocol} $\Pi$. The security of an MPC protocol is measured by its resistance to the adversary. In the semi-honest (a.k.a. honest but curious) model, the party that the adversary corrupts follow the protocols but try to learn information about the other party’s data from the transcript (\ie all messages sent and received during the protocol) of the protocol. Thus, protocols in this model are called \textit{oblivious} or \textit{private} \ie their transcripts reveal nothing beyond the protocol's results. In the malicious model, the corrupted party can also arbitrarily deviate from the established protocol arbitrarily. Thus, protocols in this model ensures any misbehavior that deviates from the protocols be detected through an additional verification scheme. A protocol that is secure against a semi-honest or malicious adversary is called a semi-honest MPC (SH-MPC) protocol $\Pi^{\text{semi}}$ or maliciously secure MPC (MS-MPC) protocol $\Pi^{\text{mali}}$, respectively. 

\partitle{Secret sharing} We use $[x]$ to denote an additive linear secret sharing (LSS) for $x$ shared between $k$ parties. Each $P_i$ holds a random share $[x]_i$ of $x$. Typically, $[x]_i$ is an element in a finite field $\field$ or a ring $\ring$, depending on the underlying MPC protocols. The shares are constructed such that their sum reconstructs the original value $x$, \ie $\sum_{i \in [k]} [x]_i = x$. We use $\share{x}$ to denote the corresponding shares in the context of MS-MPC. Each party secret-shares its plain value by invoking $\textsf{Share}$, and recovers the plain value from secret shares by invoking $\textsf{Recover}$. 
We omit the subscript and only write $[x], \share{x}$ when the ownership is irrelevant from the context.

\partitle{Malicious checking} 
To ensure a malicious security guarantee, many techniques can be used to enhance LSS with checking procedures (\eg based on Information-theoretic MACs \cite{damgaard2012multiparty,bendlin2011semi}). 
For example, $\llbracket x \rrbracket_i := ([x]_i, [\alpha \cdot x]_i)$ for a secret MAC key $\alpha$ is called \textit{authenticated share} in \cite{damgaard2013practical} that can be used to check the integrity of secret-shared values during the computation, where the corresponding checking procedure is outlined in Appendix \ref{appendix:mac_check} to provide an intuitive understanding. We remark that the checking and the form of secret-shares may vary in different MPC protocols. These techniques are well-established in public libraries (\eg MP-SPDZ \cite{mp-spdz}) to enable out-of-the-box use of function evaluation.

\partitle{Black-box functions} We use some MPC-based black-box functions in our paper, whose implementations are presented in \cite{RadixSort,mp-spdz}.

\begin{itemize}
    \item $\textsf{LT}(\cdot, \cdot)$  takes two shares $\share{x}, \share{y}$ as inputs and returns secret-shared comparison result $\share{x \overset{?}{=} y}$.
    \item $\textsf{Eq}(\cdot, \cdot)$ takes two shares $\share{x}, \share{y}$ as inputs and returns secret-shared equality $\share{x \overset{?}{=} y}$.
    \item $\textsf{ApplyPerm}(\cdot,\cdot)$ takes two vectors of shares $\share{\pi}, \share{x}$ as inputs, where $\pi$ is a permutation. It oblivious apply $\share{\pi}$ over the $\share{x}$ to obtain $\share{b}$, which means that the item $b_i$ is moved from the position $\pi(i)$ of $x$, \ie $b_i = a_{\pi(i)}$.

\end{itemize}

\subsection{Hash Function and Merkle Hash Tree}

\begin{figure}[t!]
    \centering
    \includegraphics[width=0.45\linewidth]{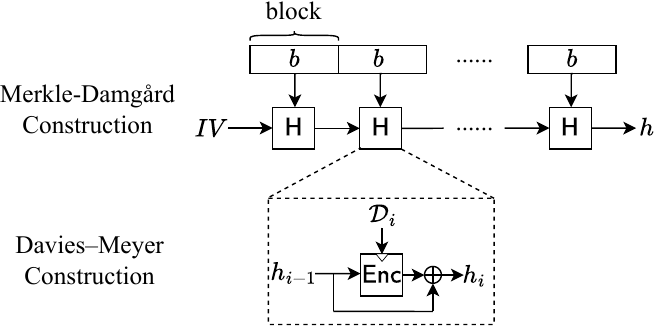}
    \vspace{-0.1in}
    \caption{Merkle-Damgård construction of a hash function instantiated by a block cipher using Davies–Meyer, where \textsf{Enc} denotes the encryption process.}
    \vspace{-0.15in}
    \label{fig:md_construction}
\end{figure}

\partitle{Merkle-Damgård construction of hash functions} The Merkle-Damgård construction is a foundational framework for cryptographic hash functions, used in standards like SHA-1/SHA-2 \cite{merkle1979secrecy}. It processes input data in fixed-length blocks using a compression function $\textsf{H}: \{0,1\}^{n} \times \{0,1\}^{b} \to \{0,1\}^{n} $. To produce the final hash, the update rule $h_i \gets \textsf{H}(h_{i-1}||\mathcal{D}_i)$ is applied in each iteration, where $h_{i-1}$ is the previous chaining value and $\mathcal{D}_i$ is the current data block. This construction guarantees collision resistance if the compression function $\textsf{H}$ is collision-resistant. This principle extends to secret-shared data by replacing $\textsf{H}$ with a corresponding MPC protocol $\Pi_\textsf{H}$. $\textsf{H}$ can be instantiated with a block cipher via Davies–Meyer Construction, as shown in Fig. \ref{fig:md_construction}. 

\noindent{\textit{Collision resistance:}} Informally, a function is collision-resistant if it is hard to find two inputs that output the same result. The formal definition of collision resistance can be found in Appendix \ref{appendix:collision_resis}. 

\partitle{Merkle hash tree} A Merkle hash tree (MHT) \cite{merkle1987digital} is a cryptographic structure that efficiently commits to a dataset while enabling succinct proofs of inclusion of an element. It operates through three key functions:

\begin{itemize}
    \item Commit: $ h_r \gets \textsf{MHT.commit}(\{h_i\}) $  
   Commits a set of elements $\{h_i\}$ by constructing an MHT and outputting the root hash $h_r$.
   \item Proof Generation: $\textsf{path}_i \gets \textsf{MHT.genProof}(i)$  
   Generates a proof for a specific element indexed $i$, consisting of sibling hashes along the path from the corresponding leaf to the root $h_r$.
   \item Verify: $ve \gets \textsf{MHT.verify}(\textsf{path}_i, h_i, h_r)$  
   A verifier recomputes the root hash using the proof $\textsf{path}_i$ and accepts the proof if the recomputed root matches $h_r$.
\end{itemize}

\begin{figure}[t!]
    \centering
    \includegraphics[width=0.3\linewidth]{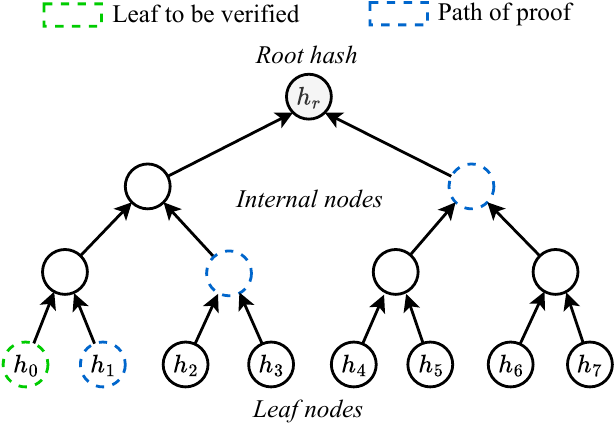}
    \vspace{-0.1in}
    \caption{A Merkle hash tree.}
    \vspace{-0.15in}
    \label{fig:merkle_tree}
\end{figure}

MHT guarantees that if a prover attempts to falsify an element $h_i'$, the recomputed hash will mismatch with $h_r$ with overwhelming probability. An illustration of MHT is shown in Fig. \ref{fig:merkle_tree}.

\subsection{Shapley Value and KNN Shapley}

The Shapley value (SV) is a common method for quantifying data contributions \cite{ghorbani2019data,jia2019efficient,agarwal2019marketplace,liu2021dealer,wang2024rethinking}. Specifically, when considering a coalition consisting of a set of contributors $I$ and the utility function $\nu$, the SV method calculates a score vector $\vec{\phi} = (sv_m)_{m\in I}$, where

\begin{equation}
sv_{m} = \frac{1}{M} \sum_{S \subseteq I \setminus m} \frac{1}{\binom{M-1}{|S|}} \left[ \nu(S \cup \{m\}) - \nu(S) \right].
\end{equation}

However, computing SV is computationally expensive \cite{shapley1953value,zhang2023efficient} and task-specific. Fortunately, recent studies \cite{jia2019efficient,wang2023threshold} show that the SV for K-Nearest Neighbors (KNN) can be computed efficiently while maintaining comparable performance to approximations for other tasks. In particular, this approach reduces the complexity from $O(2^N)$ to $O(N \log N)$ for $N$ data points. Moreover, a KNN classifier can serve as a proxy model for target learning algorithms, achieving comparable performance \cite{jia2019efficient,jia2021scalability,wang2025data} and has been applied to various domains \cite{ghorbani2022data, shim2021online, liang2020beyond, karlavs2023data,wang2025data}. However, these works assume a TCP to collect data and compute SV. The privacy risks of revealing individuals’ data to the TCP remain unaddressed \cite{wang2023threshold}.

\section{Problem Formulation And Benefits Analysis}
\label{sec:problem_formulation_benefits}
In this section, we formally define the utility signaling and its desired properties. Then, we explore its benefits for the participants. 
\subsection{Problem Formulation}
\label{sec:problem_formulation}
\subsubsection{Utility Signaling for Data Trading}
Signaling in economics refers to a method for conveying critical information between parties to facilitate informed decision-making \cite{connelly2011signaling}. In the context of data trading, we define \textit{utility signaling} as the mechanism employed to communicate the utility of a dataset offering before a transaction. Below, we provide a formal definition of utility signaling. 

\begin{definition}
(Utility Signaling) Given a selling dataset $\mathcal{D}$ from a data seller $m$, a test dataset $\mathcal{T}$ from a data buyer used to benchmark utility, and a utility evaluation function $f$, a utility signaling mechanism $\mathcal{M}$ is formally characterized as 
\begin{equation}
    \mathcal{M}: \nu \gets f(\mathcal{D}, \mathcal{T}),
\end{equation}
where $\nu$ is the utility of $\mathcal{D}$ evaluated on $\mathcal{T}$ according to $f$.  $\mathcal{M}$ determines whether $\nu$ is disclosed to the buyer, the seller, or both.
\end{definition}

The utility evaluation function $f$ is pre-agreed upon by both parties, where its output utility $\nu$ (\eg accuracy) acts as a measure of the value of the selling dataset $\mathcal{D}$. This utility can be output to the buyer to inform purchasing decisions and/or output to the seller for better pricing strategies. $\mathcal{M}$ can be abstracted as an \textit{ideal function} for computing $\nu$ and delivering it to the designated participant(s).
To realize the functionality of $\mathcal{M}$, a concrete \textit{protocol} $\Pi_{\mathcal{M}}$ can be defined to specify the operations of both participants to collaboratively implement the functionality of $\mathcal{M}$. In this work, we instantiate it with an MPC-based protocol, as detailed in \S \ref{sec:construction}.

\subsubsection{Privacy Model}
\label{sec:privacy_model} 
We assume both $\mathcal{D}$ and $\mathcal{T}$ are held privately by the seller and the buyer, respectively. $\mathcal{D}$, as the dataset intended to be sold, must remain undisclosed for economic reasons, as its disclosure would allow free duplication by the buyer and result in revenue loss. Meanwhile, $\mathcal{T}$ defines the application context and jointly determines the utility with $\mathcal{D}$, which is also often kept private due to privacy regulation or user preference. The following are examples:

\begin{itemize}
    \item In Example 1, $\mathcal{T}$ is a registered user dataset hold by Amazon. The processing and disclosure of such personal data are governed and restricted by the stringent requirements of data protection regulations like the GDPR \cite{voigt2017eu}.
    \item $\mathcal{T}$ serves as a test dataset to evaluate the accuracy of a model trained on $\mathcal{D}$. Although public test sets are commonly accessible, many practical scenarios use private test data. For instance, $\mathcal{T}$ can specify a target application (\eg clinical studies) and contains sensitive user data, which is expected to be kept confidential by the user \cite{tucker2016protecting, sim2020collaborative, zheng2023secure}.
\end{itemize}

Notably, the assumption of private $\mathcal{T}$ is mild and naturally covers the case where $\mathcal{T}$ is public, or $\nu$ does not rely on $\mathcal{T}$ ($\mathcal{T}$ can be set as an empty set).

\subsubsection{Trust Model}
\label{sec:trust_model}
We assume that both the seller and buyer are mutually distrustful and rational, each aiming to maximize their payoff. During signaling, both parties are curious to learn each other's private data due to its intrinsic value. Meanwhile, to persuade the buyer to make a purchase, the seller might manipulate the utility information conveyed to the buyer. This can be done in two ways: (1) by deviating from pre-established computations or (2) by falsifying the dataset with supposedly higher utility in the signaling process, contrary to the buyer's intention. For example, if a budget-constrained buyer plans to purchase a noisy version of a dataset \cite{liu2021dealer, agarwal2019marketplace}, the seller might input a clean version for the utility computation, misleading the buyer with an inflated signal.

\subsubsection{Desired Properties of $\Pi_{\mathcal{M}}$}
\label{sec:desired_properties}

To ensure that the output utility signal of protocol $\Pi_{\mathcal{M}}$ implement the functionality of $\mathcal{M}$, $\Pi_{\mathcal{M}}$ must satisfy the following properties:
\begin{itemize}
    \item \textbf{(Pre-purchase) Privacy}. The signaling mechanism \textit{exclusively} discloses the intended signal to designated participants, ensuring no portion of the dataset is exposed.
    \item \textbf{Reliability}. The output signal is trusted as a \textit{truthful} declaration of utility or can be efficiently verified for its truthfulness. 
\end{itemize}
 
Intuitively, a protocol $\Pi_{\mathcal{M}}$ satisfying these two properties ensures the \textit{exclusive} delivery of a \textit{truthful} signal, and can be viewed to realize the functionality of $\mathcal{M}$. The examples of cheap talk and costly signaling discussed in \S\ref{sec:intro} are special instantiations of $\Pi_{\mathcal{M}}$, where the computation of $f$ is performed locally by the seller in the formal case and by the buyer in the latter. They either guarantee privacy or ensure reliability, but not both. 

\subsection{Benefits of $\mathcal{M}$ on Data Trading}
\label{sec:application}

Before delving into the construction of $\Pi_{\mathcal{M}}$, we first demonstrate its impact on the participant decision-making and payoffs to highlight its significance. Since $\Pi_{\mathcal{M}}$ that ensures privacy and reliability can be regarded as an implementation of $\mathcal{M}$, we treat $\Pi_{\mathcal{M}}$ as a black box of $\mathcal{M}$ to explore its impact. To analyze the total utility of multiple sellers, we model them as a single seller whose dataset is the union of all individual data. 

\subsubsection{Data Trading}
\label{sec:data_trading}

A three-stage data trading via a posted price mechanism proceeds as follows:

\partitle{Signaling stage $\mathcal{M}$}
In the signaling stage, both parties collaboratively engage in the signaling mechanism $\mathcal{M}$. Specifically, the seller inputs the selling data $\mathcal{D}$, and the buyer inputs her data $\mathcal{T}$. Upon receiving $\mathcal{D}$ and $\mathcal{T}$, $\mathcal{M}$ \textit{internally} performs utility evaluation and outputs utility signal $\nu$ to participants. 

\partitle{Pricing stage $\mathcal{PR}$}
In the pricing stage, the seller posts a price $p$ based on her knowledge of the data value and the buyer, aiming to maximize her expected payoff (revenue). If the utility (value) of the data is revealed to him (\eg through $\mathcal{M}$) in this stage, the price can be adjusted based on updated knowledge.

\partitle{Deciding on purchasing stage $\mathcal{DP}$}
In this stage, the buyer decides whether to purchase, aiming to maximize her expected payoff. Upon receiving the utility signaling from $\mathcal{M}$, the buyer can update her prior knowledge of the data utility and obtain a more accurate estimate of her payoff. This helps to reduce the risk of loss of payoff and thus builds confidence in the buyer's decision.

Given that the value (utility) of data can influence the seller's pricing strategy \cite{pei2020survey}, we focus our discussion on two distinct data trading scenarios: one where signaling occurs prior to pricing, and another where it takes place afterward.

\subsubsection{Signaling Before Pricing}

\begin{figure}[t!]
    \centering
    \includegraphics[width=0.6\linewidth]{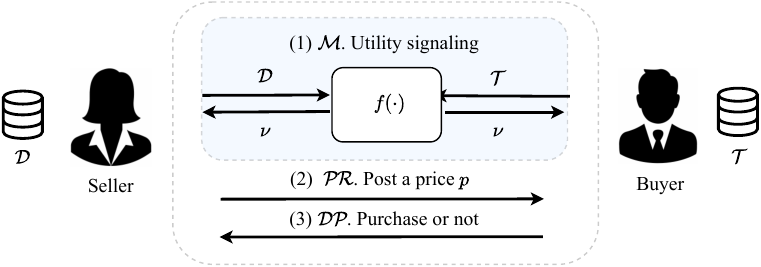}
    \vspace{-0.1in}
    \caption{Three-stage data trading where the signaling stage $\mathcal{M}$ is performed \textit{before} the pricing stage $\mathcal{PR}$. }
    \vspace{-0.15in}
    \label{fig:two_party_signaling}
\end{figure}

We first consider the scenario of signaling before pricing, where the signal $\nu$ informs both the seller's pricing and the buyer's purchasing decision. 
Then, the data trading process $\mathcal{DT}$ equipped with $\mathcal{M}$ consists of the following three stages, \ie $\mathcal{DT}_{\mathcal{M}} :=(\mathcal{M},\mathcal{PR},\mathcal{DP})$, as shown in Fig. \ref{fig:two_party_signaling}. Each party aims to maximize its payoff.

Now, we demonstrate how the utility signal $\nu$ that is outputted by $\mathcal{M}$ guides pricing and purchasing. Let $\nu$ be the utility of data, which is known to neither the seller nor the buyer at the beginning of sales process. Following \cite{chen2022selling,kamenica2011bayesian}, we assume that the seller and the buyer share the same prior belief $\lambda(\nu)$ about data utility. 
The buyer has a commonly known non-negative valuation function $u(\nu,b)$ dependent on both the data's utility $\nu$ and her private type $b$. The private type $b$ abstracts the private information the buyer holds, \eg the unit price that he would pay for one unit utility (\eg 1\% classification accuracy). We assume $b$ follows a commonly known distribution $\varphi(b)$. $u(\nu,b)$ is non-decreasing in $\nu$ and $b$. We assume $b$ and $\nu$ are continua, and $\forall b$, $u(0,b)=0$ if accuracy $\nu=0$. 

\partitle{Data trading without $\mathcal{M}$} We first consider a basic case where no signal is applied before pricing and purchasing decisions. It means data trading includes only stage 2 and stage 3, \ie $\mathcal{DT}_{-}:=(\mathcal{PR},\mathcal{DP})$. It captures various real-world data trading scenarios where decisions are conducted "blindly" and solely rely on prior knowledge. Specifically, the buyer posts a price $p$ based on nothing but her prior belief $\lambda(\nu)$.  After seeing $p$, the buyer decides to purchase the data or not based on judging whether her expected payoff is positive \ie whether her expected utility is larger than $p$. If the data is not purchased, the payoff of both parties is zero. 

Before trading, the buyer only holds $\lambda(\nu)$ to predict her payoff. Thus, the expected gain from $\mathcal{D}$ given $b$ is 
\begin{equation}
    u_0(b) \triangleq \mathbb{E}_\nu[u(\nu,b)] = \int_{\nu} u(\nu,b)\lambda (\nu) \text{d}\nu.
\end{equation}

Given the data price $p$, the expected payoff of the buyer is 
\begin{equation}
\label{equ:buyer_basic_revenue}
\mathcal{P}_{buyer,\mathcal{DT}_{-}} (p) = u_0(b) - p =  \int_{\nu} u(\nu,b)\lambda (\nu) \text{d}\nu - p.
\end{equation}
For the seller, the payoff depends on whether the buyer purchases or not \ie whether $\mathcal{P}_{buyer,\mathcal{DT}_{-}}$ is positive. Note that $b$ is private to the buyer, the seller must decide on her prior belief  $\varphi(b)$. Let $\mathbbm{1}(\cdot)$ be an indicator function, the expected payoff of the seller is 
\begin{equation}
\label{equ:seller_basic_revenue}
    \mathcal{P}_{seller,\mathcal{DT}_{-}} (p)= p \cdot \int_b \mathbbm{1}(\mathcal{P}_{buyer,\mathcal{DT}_{-}} > 0)\varphi(b) \text{d}b.
\end{equation}

We see that the seller only obtains positive payoff when the buyer decides to purchase \ie $p < u_0(b) $ given $b$. In fact, since $b$ follows a probability density function $\varphi(b)$, the continuous distribution of $u_0(b)$ can be derived based on this, denoted as $F_0$ (thus $F_0(p) = \Pr[u_0(b)<p]$.), with its corresponding density function $f_0$. Consequently, equation \eqref{equ:seller_basic_revenue} can be reformulated as

\begin{equation}
\label{equ:seller_basic_revenue_2}
    \mathcal{P}_{seller,\mathcal{DT}_{-}} (p) = p \cdot (1 - F_0(p))
\end{equation}

Thus, the seller can find  $p^*_{\mathcal{DT}_{-}}= \text{argmax}_p \mathcal{P}_{seller,\mathcal{DT}_{-}}(p)$ to maximize her expected payoff. This optimization problem is very common in economic literature \cite{Myerson1981OptimalAD, Perloff1984EquilibriumWP} and typically requires the mild technical assumption that $1 - F_0$ exhibits log-concavity – \ie the composition $\log(1 - F_0)$ forms a concave function. This property is satisfied by most canonical probability distributions, including but not limited to the uniform, exponential, and normal distributions.

\begin{lemma}
\label{lem:unique_solution}
    The seller's payoff maximization problem \eqref{equ:seller_basic_revenue_2} has a unique solution if $1 - F_0$ is log-concave.
\end{lemma}

The proof is in Appendix \ref{appendix:proof_lemma1}.

Denote the unique solution of problem \eqref{equ:seller_basic_revenue_2} as $p_0^*$. After seeing $p^*_0$, the buyer decides whether to purchase by determining $\mathcal{P}_{buyer,\mathcal{DT}_{-}} (p^*_0)> 0$. Let $\mathcal{P}^{true}$ be the true payoff (ex post payoff) of participants \textit{after} data transaction. The buyer can actually obtain the data if the buyer decides to purchase and her true payoff is $\mathcal{P}_{buyer,\mathcal{DT}_{-}}^{true} = u(\nu, b) - p^*_0$. At the same time, $\mathcal{P}_{seller,\mathcal{DT}_{-}}^{true} = p^*_0$.

\partitle{Data trading with ${\mathcal{M}}$} Now we analyze $\mathcal{DT}_{\mathcal{M}}$ and compare it with $\mathcal{DT}_{\mathcal{-}}$. $\mathcal{DT}_{\mathcal{M}}$ allows both parties to obtain the utility $\nu$  before pricing and purchasing. Let $p$ be the posted price from the seller. Then, the buyer's payoff can be obtained directly based on the valuation function $u(\nu, b)$. Because $\nu$ is a deterministic value, we can neglect $\nu$, denoting $u(b) \triangleq u(\nu, b)$. So the buyer's payoff is
\begin{equation}
\mathcal{P}_{buyer,\mathcal{DT}_{\mathcal{M}}} (p) = u(b) - p.
\end{equation}

Similar to equations (\ref{equ:buyer_basic_revenue}), (\ref{equ:seller_basic_revenue}) and (\ref{equ:seller_basic_revenue_2}), the continuous distribution of $u(b)$ can be derived based on $\varphi(b)$, denoted as $F$, with its corresponding density function $f$. Thus, we have

\begin{equation}
\label{equ:seller_pis_revenue_2}
    \mathcal{P}_{seller,\mathcal{DT}_{\mathcal{M}}}(p) = p \cdot (1 - F(p)).
\end{equation}

Similar to Lemma \ref{lem:unique_solution}, under the mild assumption that $1 - F$ is log-concave, the optimal solution $p^*= \text{argmax}_p \mathcal{P}_{seller,\mathcal{DT}_{\mathcal{M}}}(p)$ is unique. 

To demonstrate the superiority of $\mathcal{DT_{\mathcal{M}}}$, we aim to prove that compared with the basic case $\mathcal{DT_{-}}$, $\mathcal{DT_{\mathcal{M}}}$ enables both the data seller and buyer to achieve enhanced satisfaction with their respective payoff (utility) in the data transaction. Specifically: 
\begin{enumerate}
    \item When the realized data quality exceeds prior expectations (\ie $u(b) > u_0(b)$), the seller should obtain higher ture payoff (\ie $\mathcal{P}_{seller,\mathcal{DT}_{\mathcal{M}}}^{true} > \mathcal{P}_{seller,\mathcal{DT}_{-}}^{true}$); otherwise, the seller would perceive themselves as "selling at a loss" (suboptimal transaction pricing);
    \item When the realized data quality falls below prior estimates (\ie $u(b) < u_0(b)$), $\mathcal{DT_{\mathcal{M}}}$ induces some buyer withdrawal from unworthy transactions, thereby preventing "buyer's remorse" (suboptimal purchases) thus improving buyer's true utility (\ie $\mathcal{P}_{buyer,\mathcal{DT}_{\mathcal{M}}}^{true} > \mathcal{P}_{buyer,\mathcal{DT}_{-}}^{true}$).
\end{enumerate}

The proof of statement (1) follows standard arguments, whereas the proof of statement (2) requires an additional mild technical condition. We first formalize and prove statement (1) as follows.

\begin{theorem}
\label{thm:seller_rev_inc}
    If $u(b) > u_0(b)$, then 
    $$\mathcal{P}_{seller,\mathcal{DT}_{\mathcal{M}}}^{true}(p^*) > \mathcal{P}_{seller,\mathcal{DT}_{-}}^{true}(p_0^*).$$
\end{theorem}

Following the methodology in the proof of \autoref{thm:seller_rev_inc} (in Appendix \ref{appendix:proof_theorem1}), consider the scenario where $u(b) < u_0(b)$. This induces first-order stochastic dominance (FOSD) with $F(p) > F_0(p)$ for all $p$. However, mere FOSD is insufficient to establish statement (2). We therefore strengthen our analytical framework by adopting the more potent hazard rate dominance criterion:

\begin{definition} \textup{\cite{krishna2009auction}}
    $F$ is hazard rate dominated by $F_0$ if  
    \[
        \frac{1 - F(p)}{f(p)} < \frac{1 - F_0(p)}{f_0(p)}, \quad \forall p \in \text{supp}(F) \cap \text{supp}(F_0),
    \]
    where $\text{supp}(F)$ denotes the support of distribution $F$.
\end{definition}

This condition implies but is not implied by FOSD. Through this refined dominance relationship, we derive the next lemma which states that $\mathcal{DT_{\mathcal{M}}}$ yields strictly lower data price than $\mathcal{DT_{-}}$.

\begin{lemma}
\label{lem:price_decrease}
    Assume $1 - F_0$ and $1 - F$ are log-concave, and $F$ is hazard rate dominated by $F_0$, if $u(b) < u_0(b)$, then $p^* < p^*_0$.
\end{lemma}

The proof is in Appendix \ref{appendix:proof_lemma2}. Then we can prove the statement (2) based on this lemma:

\begin{theorem}
\label{thm:buyer_uti_inc}
    If $u(b) < u_0(b)$, then 
    
    $$\mathcal{P}_{buyer,\mathcal{DT}_{\mathcal{M}}}^{true}(p^*) > \mathcal{P}_{buyer,\mathcal{DT}_{-}}^{true}(p^*_0)$$
    
    \noindent under the assumption of Lemma \ref{lem:price_decrease}.
\end{theorem}

The proof is in Appendix \ref{appendix:proof_theorem2}. 

\begin{figure}[t!]
    \centering
    \begin{subfigure}[b]{0.25\textwidth} 
        \centering
        \includegraphics[width=\linewidth]{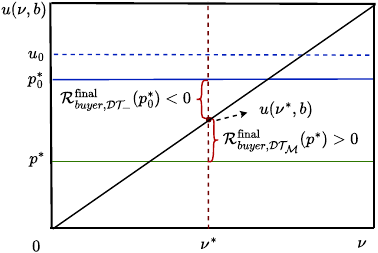} 
        \caption{$u_0(b) \geq p_0^*, u(b) \geq p^*$.}
        \label{fig:revenue_case1}
    \end{subfigure}
    \hspace{0.1in}
    \begin{subfigure}[b]{0.25\textwidth} 
        \centering
        \includegraphics[width=\linewidth]{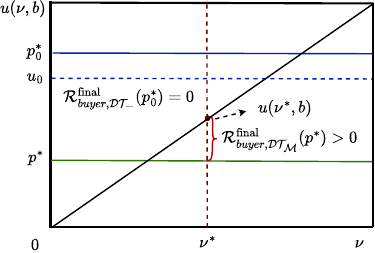} 
        \caption{$u_0(b) < p_0^*, u(b) \geq p^*$.}
        \label{fig:revenue_case2}
    \end{subfigure}
    \vspace{-0.1in}
    \caption{Intuitive examples of the \autoref{thm:buyer_uti_inc}.}
    \vspace{-0.2in}
    \label{fig:examples}
\end{figure}

\partitle{Intuitive examples} To intuitively illustrate the result of \autoref{thm:buyer_uti_inc}, we give two intuitive examples as shown in Fig. \ref{fig:examples}. In Fig. \ref{fig:revenue_case1}, though the true payoff of purchasing is negative ($u(\nu^*,b) < p_0^*$), the buyer in $\mathcal{DT}_{-}$ has overestimate expectation ($u_0 > p_0^*$) that drives him into loss-incurring purchase (loss $p^*_0 -u(\nu^*,b)$ payoff). While in $\mathcal{DT}_{\mathcal{M}}$ with a signaling stage, $\mathcal{M}$ delivers the utility $\nu$ to the both parties. After knowing the utility is lower than her expectation, the seller reduces the price to $p^*$ to encourage purchasing. This updated price, however, could result in a positive payoff ($u(\nu^*,b) - p^*$) for the buyer, thereby facilitating a successful trade. A similar insight can be observed in Fig. \ref{fig:revenue_case2}, where signaling transforms a failed transaction into a successful one. Another concrete example of $\mathcal{M}$ protects the buyer's payoff is detailed in Appendix \ref{appendex:payoff_example}.

\partitle{Discussion on the case where $b$ is public}
Consider the scenario in which the buyer publicly reveals her type $b$ to the seller\footnote{The revealed $b$ may not be the true type of the seller \cite{agarwal2019marketplace}. Nevertheless, it still specifies the acceptable threshold of $p$ on $\nu$ of the buyer.}. We denote this case as $\mathcal{DT}_{\mathcal{M}}^b$, and have the following observation. The details is illustrated in Appendix \ref{appendix:public_b}.

\begin{theorem}
    \label{thm:signaling_b_public}
    $\mathcal{P}_{seller,\mathcal{DT}_{\mathcal{M}}^b}^{\text{true}} \ge 0, \mathcal{P}_{buyer,\mathcal{DT}_{\mathcal{M}}^b}^{\text{true}} \ge 0$.
\end{theorem}

\subsubsection{Signaling After Pricing}

Next, consider the scenario where signaling is conducted after pricing \ie $\mathcal{DT}_{\mathcal{M}}' :=(\mathcal{PR},\mathcal{M}, \mathcal{DP})$.
In this case, the pricing $p$ is fixed in advance, and the signaling influences only the decision of the buyer. 

Given the signal $\nu$, the buyer can now effectively refuse to purchase for $\nu$ that $ u(\nu,b) - p < 0 $ to ensure her true payoff is non-negative. We have the following theorem.

\begin{theorem}
    \label{thm:signaling_after_pricing}
    If $p$ is fixed and $\nu$ is given to the buyer, then 
    $$\mathcal{P}_{buyer,\mathcal{DT}_{\mathcal{M}}'}^{\text{true}} \ge \mathcal{P}_{buyer,\mathcal{DT}_{\mathcal{-}}'}^{\text{true}}.$$
\end{theorem}

The proof is in Appendix \ref{appendix:proof_theorem3}.

\subsubsection{Summary of Results}
Our above analysis demonstrates several key results. When both participants access the utility signal $\nu$ pre-transaction, \autoref{thm:seller_rev_inc} and \autoref{thm:buyer_uti_inc} show $\mathcal{M}$ can effectively prevent suboptimal decisions for both participants when the actual utility deviates from expectation, no matter $b$ is private or public. In the case where $\nu$ can only be accessed by the buyer, \autoref{thm:signaling_after_pricing} shows that the buyer's expected true payoff increases immediately. Crucially, in all scenarios, the buyer's true payoff is guaranteed to be non-negative.

These results demonstrate that $\mathcal{M}$ incentivizes risk-averse buyers by eliminating the risk of financial loss, thus facilitating transactions that might not otherwise occur. Furthermore, when the seller also has access to the utility before the pricing stage, $\mathcal{M}$ opens the possibility to improve her payoff by adjusting prices for high-value data. Since many buyers in the real world are of risk-averse type \cite{markowits1952portfolio}, $\mathcal{M}$ essentially provides a way for them to strictly prevent suffering payoff loss. This encourages them to participate in data trading, thus helping to improve social welfare. 

\section{Construction of $\Pi_{\mathcal{M}}$}
\label{sec:construction}
Having established the theoretical advantages of signaling, we now turn to its concrete realization. In this section, we detail the construction of a utility signaling protocol $\Pi_{\mathcal{M}}$ that implements the functionality of $\mathcal{M}$ and meets the crucial requirements of reliability and privacy as outlined in \S\ref{sec:desired_properties}.

\subsection{Basic Protocol}

We first illustrate a straw-man protocol $\Pi_{\mathcal{M}}^{\text{semi}}$ to ensure privacy, which is straightforward but suffers from reliability vulnerability. Below, we define $\Pi_{f}$, the MPC-based protocol for utility calculation $f$, as a foundation for introducing $\Pi_{\mathcal{M}}^{\text{semi}}$.

\begin{definition}
    Given secret-shared input dataset $\share{\mathcal{D}}$ and $\share{\mathcal{T}}$, the MPC-based utility calculation protocol is defined as:
    \begin{equation}
        \share{\nu} \gets \Pi_{f}(\share{\mathcal{D}}, \share{\mathcal{T}})
    \end{equation}
\end{definition}


Since the utility calculation $f$ is task-dependent and can be decomposed as basic modules (\eg scalar multiplication and sorting) whose MPC-based implementation has been extensively studied and well-supported by literature \cite{mp-spdz,emp-toolkit},  
we treat $\Pi_{f}$ as a 
\textit{black-box} MPC protocol and do not delve into its specific instantiation. 

Next, we construct $\Pi_{\mathcal{M}}^{\text{semi}}$, a utility signaling mechanism that ensures privacy, by leveraging SH-MPC modules: $\textsf{Share}^{\text{semi}}$, $\textsf{Recover}^{\text{semi}}$, and $\Pi_{f}^{\text{semi}}$. 
The process is detailed in Alg. \ref{alg:signaling_semi}. First, the dataset is transformed into a secret-shared form to enable MPC-based computation (line 1). Next, the parties collaboratively execute the protocol $\Pi_{\nu}^{\text{semi}}$ to compute a secret-shared utility signal $[\nu]$ (line 2). Subsequently, the parties jointly recover $[\nu]$ to its plaintext form $\nu$ (line 3), completing the signaling process. The security guarantees of SH-MPC ensures $\Pi_{\mathcal{M}}^{\text{semi}}$ satisfies privacy.

\begin{algorithm}\small
\caption{$\Pi_{\mathcal{M}}^{\text{semi}}$}
\label{alg:signaling_semi}
\KwIn{Selling data $\mathcal{D}$, test data $\mathcal{T}$}
\KwOut{The utility signal $\nu$}
$[\mathcal{D}] \gets \textsf{Share}^{\text{semi}}(\mathcal{D})$, $[\mathcal{T}] \gets \textsf{Share}^{\text{semi}}(\mathcal{T})$ \hfill  $\triangleright$ secret share

$[\nu] \gets \Pi_{f}^{\text{semi}}([\mathcal{D}], [\mathcal{T}])$

$\nu \gets \textsf{Recover}^{\text{semi}}([\nu])$ \hfill  $\triangleright$ recover to plaintext

\Return{$\nu$}
\end{algorithm}

\partitle{Reliability attacks}  
We demonstrate two simple yet effective attacks on $\Pi_{\mathcal{M}}^{\text{semi}}$ that compromise its reliability. (1) \textit{Deviation of Computation}. The first attack exploits deviations in the computation process.
An attacker (\eg the seller) aiming to manipulate the utility result can locally tamper with the secret-shared utility signal $[\nu]$ by introducing an offset $\epsilon$, yielding $\hat{[\nu]} = [\nu] + \epsilon$. When this manipulated value is input into $\textsf{Recover}^{\text{semi}}$, the output $\nu$ becomes $\nu + \epsilon$, undetected by the other party. This allows the attacker to arbitrarily inflate or deflate $\nu$ to suit their interests (\eg, an inflated $\nu$ could mislead the buyer into purchasing the data, thereby increasing the seller's payoff). Notably, such manipulation can also occur within other sub-modules, such as $\Pi_{\nu}^{\text{semi}}$. (2) \textit{Falsification of Input}. The second attack is even simpler: the seller can input falsified data $\hat{\mathcal{D}}$ into $\Pi_{\mathcal{M}}^{\text{semi}}$ to manipulate $\nu$ without detection by the buyer.  

These vulnerabilities render $\Pi_{\mathcal{M}}^{\text{semi}}$ unreliable, as the truthfulness of the output $\nu$ cannot be guaranteed. In the following, we demonstrate how to enhance $\Pi_{\mathcal{M}}^{\text{semi}}$ to achieve reliability.

\vspace{-0in}
\subsection{Ensuring Reliability in Utility Signaling}

The primary objective in ensuring signal reliability is to address potential vulnerabilities in the signaling process that could undermine the integrity of the computed results. As highlighted in \S\ref{sec:trust_model} and by the two aforementioned attacks, the reliability of $\Pi_{\mathcal{M}}$ can be compromised by two misbehavior: (1) deviations in the computation process of $\Pi_{\mathcal{M}}$, and (2) falsification of the input data $\mathcal{D}$. Accordingly, the reliability goal can be decomposed into two sub-goals, each aimed at mitigating these specific misbehaviors.
\begin{enumerate}
\item \textbf{Robustness of Computation (RoC)}: Any deviation in the computational process of the signal must be detected with high probability.
\item \textbf{Authenticity of Inputs (AoI)}: Any falsification of the data $\mathcal{D}$ input to the signal computation must be detected with high probability.
\end{enumerate}

Together, these two properties guarantee that the signal correctly reflects the utility of the data and that no adversarial deviation or manipulation occurs during computation. Next, we demonstrate $\Pi_\mathcal{M}^{\text{mali}}$ that can achieve both the desired properties.

\subsubsection{Overview}

\begin{figure}[t!]
    \centering
    \includegraphics[width=0.55\linewidth]{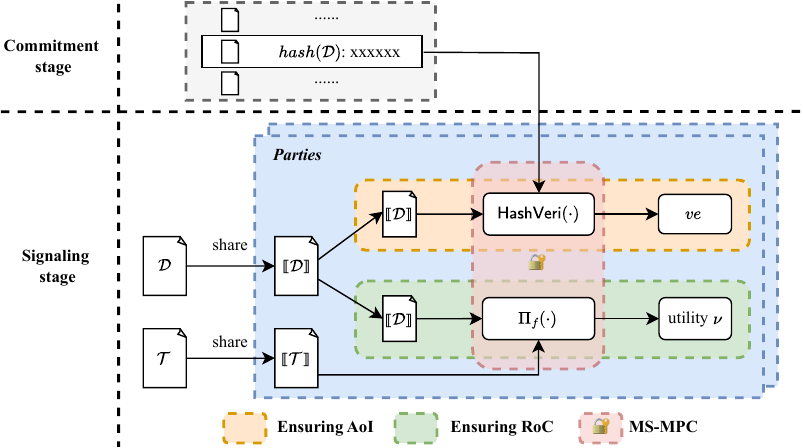}
    \vspace{-0.1in}
    \caption{High-level ideas of ensuring the reliability of signal.}
    \vspace{-0.15in}
    \label{fig:high_level_ideas}
\end{figure}

We present $\Pi_\mathcal{M}^{\text{mali}}$ which can achieve the reliability requirements of RoC and AoI, simultaneously. At a high level, $\Pi_\mathcal{M}^{\text{mali}}$ achieves them by making both computation and input verifiable by integrating cryptographic verification methods. The former is done by improving the utility calculation with MS-MPC, and the latter is achieved by incorporating an MS-MPC-based hash verification protocol with a commitment stage.  We outline $\Pi_\mathcal{M}^{\text{mali}}$ in Alg. \ref{alg:full_mechanism}. It takes as input the selling data $\mathcal{D}$, test data $\mathcal{T}$, a committed hash value $h$, and a block size $b$. It outputs a utility value $\nu$, or aborts with failure when the hash verification fails. A graphic workflow of $\Pi_\mathcal{M}^{\text{mali}}$ is depicted in Fig. \ref{fig:high_level_ideas}. Specifically, $\Pi_\mathcal{M}^{\text{mali}}$ improves $\Pi_\mathcal{M}^{\text{semi}}$ in the following aspects. 

\begin{algorithm}\small
\caption{$\Pi_{\mathcal{M}}^{\text{mali}}$}
\label{alg:full_mechanism}
\KwIn{Selling data $\mathcal{D}$, test data $\mathcal{T}$, committed hash value $h$, block size $b$}
\KwOut{The utility signal $\nu$}
$\share{\mathcal{D}} \gets \textsf{Share}^{\text{mali}}(\mathcal{D})$, $\share{\mathcal{T}} \gets \textsf{Share}^{\text{mali}}(\mathcal{T})$ 

$ve = \textsf{HashVeri}^{\text{mali}}(\share{\mathcal{D}}, h,b)$

\If{$!ve$} {
    parties abort
}

$\share{\nu} \gets \Pi_{f}^{\text{mali}}(\share{\mathcal{D}}, \share{\mathcal{T}})$

$\nu \gets \textsf{Recover}^{\text{mali}}(\share{\nu})$ 

\Return{$\nu$}
\end{algorithm}

\noindent\partitle{Ensuring RoC} 
$\Pi_\mathcal{M}^{\text{mali}}$ replaces the modules $\textsf{Share}^{\text{semi}}$, $\textsf{Recover}^{\text{semi}}$, and $\Pi_{f}^{\text{semi}}$ in $\Pi_\mathcal{M}^{\text{semi}}$ with their MS-MPC version (lines 1, 5, 6) to ensure RoC. Recall that MS-MPC primitives inherently ensure the integrity of computing progress through embedded verification mechanisms. Thus, this improvement directly prevents any parties from deviating from the computation since any attempt to manipulate the results during the computation will be directly detected. We provide a concrete example of a checking mechanism based on \cite{damgaard2013practical} (detailed in Appendix \ref{appendix:mac_check}) to intuitively illustrate how detection works\footnote{For simplicity, this example illustrates the check for a single computed value. The full protocol in Appendix \ref{appendix:mac_check} shows how to efficiently check multiple values in a batch using a random linear combination.}.

\begin{example}
Consider two parties, $P_0$ and $P_1$, computing the sum $z=x+y$. Each party $P_i$ holds authenticated shares, denoted as $\llbracket v \rrbracket_i = ([v]_i, [\gamma(v)]_i)$, where $[v]$ is the additive shares of $x$ or $y$, $[\gamma(v)]_i$ is the $i$-th share of the MAC $\gamma(v) = \alpha \cdot v$, $\alpha$ is the secret MAC key. After correctly computing shares for the sum, $[z]_i$, and its MAC, $[\gamma(z)]_i$, suppose the malicious party $P_1$ tampers with its value share to create $[z]'_1 = [z]_1 + \epsilon$. When the parties open the value, the reconstructed public result becomes $\hat{z} = z + \epsilon$. To verify this result, they perform a MAC check by having each party $P_i$ compute a share of a check value, $[\sigma]_i := [\gamma(z)]_i - \hat{z} \cdot [\alpha]_i$. Upon reconstruction, the final check value is: $\sigma = \sum_i ([\gamma(z)]_i - \hat{z} \cdot [\alpha]_i) = \gamma(z) - \hat{z} \cdot \alpha = -\alpha\epsilon$. Since both the secret MAC key $\alpha$ and the manipulation offset $\epsilon$ are non-zero, $\sigma$ will be non-zero and leads to check failure. \cite{damgaard2013practical} ensures that any deviation from the prescribed computation will lead to a check failure with overwhelming probability.
\end{example}

\noindent\partitle{Ensuring AoI} $\Pi_\mathcal{M}^{\text{mali}}$ includes a hash verification procedure (lines 2-4) with a commitment stage to ensure AoI. To establish the correctness of input data $\mathcal{D}$, $\Pi_\mathcal{M}^{\text{mali}}$ requires a pre-committed $h$, which is the hash of the committed data. This reduces the problem of ensuring the correctness of inputs to verify the consistency of input between the commitment stage and signaling stage. By computing the hash from $\share{\mathcal{D}}$ and comparing it with the committed hash (line 2), the consistency is verified due to the collision-resistance nature of a hash function. A verification failure breaks the correctness of input $\mathcal{D}$, and parties will abort the protocol (lines 3-4).

\begin{remark}
(Why does the collision resistance build the consistency?) At the core of hash verification is the collision resistance feature of cryptographic hash functions. This property makes it computationally hard to find two distinct inputs that produce the same hash (see Appendix \ref{appendix:collision_resis} for details). This ensures that a buyer cannot forge a different dataset $\hat{\mathcal{D}}$ that matches the committed hash of $\mathcal{D}$, as doing so would require breaking collision resistance.
\end{remark}

\noindent\partitle{The commitment stage} To establish a foundation for AoI, a commitment stage is necessary prior to the signaling stage to define the correctness of the input data. During this stage, the seller publishes a hash value $h = \textsf{commit}(\mathcal{D})$ in a public bulletin (\eg a well-established website or blockchain). This hash acts as a commitment to the dataset, allowing further verification to ensure that subsequent signal computation is tied to the committed data.

\begin{remark}
(Why does a commitment stage establish the correctness of data?) The commitments published in a public bulletin are publicly verifiable and historically trackable by all buyers and potential auditors. As such, the correctness of it can be assumed to be verified, as any mismatch between data and the corresponding commitment in data trading undermines both the seller's payoff and reputation.
\end{remark}

Therefore, given the correctness of the long-term commitment, once the consistency between the committed data and the temporary input $\mathcal{D}$ of signaling is confirmed, the correctness of $\mathcal{D}$ (AoI) is directly guaranteed. Next, we detail how to establish the input correctness by incorporating hashing and commitment.

\subsubsection{Hash Verification Protocol}
\label{sec:hash_ver_pto}
At the core of $\Pi_\mathcal{M}^{\text{mali}}$ is the hash verification protocol $\textsf{HashVeri}^{\text{mali}}$ to build the correctness of input. We propose two protocols to trade off between the rigor of AoI guarantees and efficiency.

\partitle{Merkle-Damgård-based hash verification protocol}
The first solution is the Merkle-Damgård-based hash verification protocol that achieves a strict AoI guarantee. The idea behind it is simple: input correctness is established if the plaintext hash evaluation during the commitment stage matches the MPC-based hash evaluation result. Accordingly, we first describe the hash computation process and then illustrate the complete verification protocol. 

\textit{Hash evaluation}. To rigorously ensure that any falsification of data will be detected, the hash computation must process \textit{all} data blocks. 
For succinct hash output (\eg 128 bits) from potentially large data, we employ Merkle-Damgård hash construction \cite{merkle1979secrecy}, leveraging a compression function $\textsf{H}$ to compute the hash in a chaining manner. The compression function $\textsf{H}$ can be instantiated with standard hash circuits, such as the SHA-512 circuit. In the commitment stage, the hash is computed on plaintext, \ie compute $h = \textsf{commit}(\mathcal{D})$. Then, the hash is evaluated using MPC primitives in the signaling stage, acting as a proof of input correctness. 

\textit{Hash verification protocol $\textsf{HashVeri}_{\textsf{MD}}^{\text{mali}}$}.
In the signaling stage, we propose hash verification protocol $\textsf{HashVeri}_{\textsf{MD}}^{\text{mali}}$ to ensure the correctness of input, as detailed in Alg. \ref{alg:hash_veri_md}. The key step involves evaluating the hash of secret-shared input $\mathcal{\mathcal{D}}$ via the Merkle-Damgård construction (line 2-5) and comparing it with the commitment (line 6). Due to the collision-resistant property of a cryptographic hash, any falsification of $\share{\mathcal{D}}$ leads to verification failure with overwhelming possibility. This establishes the consistency of $\share{\mathcal{D}}$ with the committed $\mathcal{D}$. All the operations are performed with MS-MPC primitives to guarantee the integrity of the computation.

\begin{algorithm}\small
\caption{$\textsf{HashVeri}_{\textsf{MD}}^{\text{mali}}$}
\label{alg:hash_veri_md}
\KwIn{Secret shared data $\share{\mathcal{D}}$, a hash value $h$, block size $b$}
\KwOut{Verification result $ve \in \{0,1\}$}
$\{\share{\mathcal{D}_i}\}_{i\in \{1,...,k\}} \gets \textsf{split}_{b}(\share{\mathcal{D}})$

$\share{h_0} \gets 0^n$ \hfill  $\triangleright$ initialization vector

\For{$i \gets 1$ \KwTo $k$}{ 
    $\share{h_i} \gets \Pi_{\textsf{H}}^{\text{mali}}(\share{h_{i-1}}||\share{\mathcal{D}_i})$ \hfill  $\triangleright$ Merkle-Damgård
}

$h_k \gets \textsf{Recover}^{\text{mali}}(\share{h_k})$

(Buyer) $ve \gets h\overset{?}{=} h_{k}$

\Return{$ve$}
\end{algorithm}

\partitle{Approximate hash verification protocol} While the Merkle-Damgård-based approach ensures a strict AoI guarantee, its overhead scales linearly with the number of data blocks. To improve its efficiency, we propose an approximate hash verification scheme that reduces the number of evaluated blocks by random sampling. The idea is inspired by data integrity checks of remote cloud data storage \cite{ateniese2007provable, burke2025scalable}, where rigorous verification incurs large data transmission and can be addressed by sampling with a proper data structure.

\begin{definition}
    ($(\alpha,\beta)$-AoI) Given seller falsifies $\alpha$ ratio data blocks, the $\textsf{HashVeri}$ output $False$ with at least $\beta$ probability.
\end{definition}

To commit the data $\mathcal{D}$, the buyer first splits it into blocks and computes their individual hashes, yielding $\{\textsf{H}(\mathcal{D}_i)\}$. Instead of using a Merkle–Damgård chaining approach, we construct an MHT, where each $\{\textsf{H}(\mathcal{D}_i)\}$ serves as a leaf node of the tree, and the root hash $h_r = \textsf{MHT.commit}(\{\textsf{H}(\mathcal{D}_i)\})$ represents the commitment. 

The hash verification process in the signaling stage is detailed in Alg. \ref{alg:hash_veri_ap}. To ensure AoI, the buyer randomly samples a set of indices $I$ of size $c$, representing the challenged block indices (line 2). The parties then collaboratively compute the hash of these blocks $\{h_i\}_{i \in {I}}$, using the MS-MPC hash evaluation $\Pi_{\textsf{H}}^{\text{mali}}(\{\share{\mathcal{D}_i}\}_{i \in I})$ (lines 3-5). To prove $\{h_i\}_{i \in {I}}$ are indeed included as leaf nodes in the committed MHT rooted at $h_r$, the seller additionally provides $\{\textsf{path}_i\}_{i \in I}$ where each $\textsf{path}_i$ contains all siblings of the nodes along the path from the $i$-th leaf to the root $h_r$ (line 6). By verifying the consistency between $\{h_i\}_{i \in {I}}$ and $h_r$ through $\{\textsf{path}_i\}_{i \in I}$ (line 9), the buyer ensures the correctness of the $c$ sampled blocks.

Due to the collision-resistant property of hash and the security guarantee of MHT, the integrity of challenged blocks is inherently ensured. This approach significantly reduces the MPC evaluation overhead by performing fewer $\Pi_{\textsf{H}}^{\text{mali}}$ evaluations on blocks. As a trade-off, there exists the possibility that the seller's falsified blocks are not detected, which depends on the number of falsified blocks and challenged blocks. We demonstrate $\textsf{HashVeri}_{\textsf{AP}}^{\text{mali}}$ achieves a $(\frac{t}{k},1-(\frac{k-t}{t})^c)$ guarantee of AoI.

\begin{algorithm}\small
\caption{$\textsf{HashVeri}_{\textsf{AP}}^{\text{mali}}$}
\label{alg:hash_veri_ap}
\KwIn{Secret shared data $\share{\mathcal{D}}$, a root hash $h_r$ of a MHT, block size $b$}
\KwOut{Verification result $ve \in \{0,1\}$}
$\{\share{\mathcal{D}_i}\}_{i\in \{1,...,k\}} \gets \textsf{split}_{b}(\share{\mathcal{D}})$

(Buyer) $I \gets \{{i|i \overset{\textsf{rd}}{\gets} \{1,...,k\} \}}$ where $|I|=c$, send $I$ to the seller

\For{$i \in I$}{ 
    $\share{h_i} \gets \Pi_{\textsf{H}}^{\text{mali}}(\share{\mathcal{D}_i})$ 
    
    $h_i \gets \textsf{Recover}^{\text{mali}}(\share{h_i})$
}

(Seller) send $\{\textsf{path}_i\}_{i \in I}$ to the buyer, where $\textsf{path}_i \gets \textsf{MHT.genProof}(i)$

(Buyer) $ve \gets False $

\For{$i \in I$}{ 
    $ve \gets ve \And \textsf{MHT.verify}(\textsf{path}_i, h_i, h_r) $
}

\Return{$ve$}
\end{algorithm}

\begin{theorem}
    \label{thm:app_hash_veri}
    Given total number of data blocks be $k$, number of falsified blocks be $t$, number of challenged blocks be $c$, $\textsf{HashVeri}_{\textsf{AP}}$ achieves $(\frac{t}{k},1-(\frac{k-t}{k})^c)$-AoI.
\end{theorem}

The proof is in Appendix \ref{appendix:proof_theorem4}.

\partitle{Instantiation of $\textsf{H}$} A natural choice for $\mathsf{H}$ is a popular hash function, such as SHA-512, where $\textsf{H} : \{0, 1\}^{1536} \rightarrow \{0, 1\}^{512}$. However, the cost of evaluating $\textsf{H}$ with MPC is dominated on the number of the $\textsf{AND}$ gates in the whole circuit ($\textsf{XOR}$ gates can be evaluated locally with no communications). For instance, using SHA-512 would require 57,947 $\textsf{AND}$ gates to commit a 1024-bit message\footnote{https://nigelsmart.github.io/MPC-Circuits/}, averaging to 56 $\textsf{AND}$ gates per bit. To address this inefficiency, we leverage LowMC \cite{albrecht2015ciphers}, an MPC-friendly block cipher, to instantiate $\textsf{H}$ by using Davies–Meyer construction \cite{katz2007introduction} as below:
\begin{equation}
    \textsf{H}:h_{i} \gets \textsf{LowMC.Enc}(\mathcal{D}_i,h_{i-1}) \oplus h_{i-1}
\end{equation}
Compared to standard block ciphers like AES (averaging 50 $\textsf{AND}$ gates per bit), LowMC with 128-bit key and block size significantly reduces the number of $\textsf{AND}$ gates (averaging 4.6 $\textsf{AND}$ gates per bit). 

\partitle{Efficiency Analysis} For $\textsf{HashVeri}_{\textsf{MD}}^{\text{mali}}$, the complexity is dominated by the number of $\textsf{AND}$ gates in the evaluated circuit of the compression function $\textsf{H}$. Given $n$ as data size, $b$ as compression block size, $k$ as the number of $\textsf{AND}$ gates in $\textsf{H}$. The total number of $\textsf{AND}$ gates is $k{\lceil n/b \rceil}$. For $\textsf{HashVeri}_{\textsf{AP}}^{\text{mali}}$, $(\alpha, \beta)$-AoI requires the number of MPC-evaluated challenged blocks to be a constant $c = \frac{\ln(1-\beta)}{\ln (1-\alpha)}$ based on Theorem \ref{thm:app_hash_veri}, so the total number of $\textsf{AND}$ gates is $\frac{k\ln(1-\beta)}{\ln (1-\alpha)}$. The MHT proof generation and verification are performed on plain data and enjoy complexity of $O(\log(\lceil n/b \rceil))$.

\subsubsection{Discussion on the Promised Guarantees} (Sketch) The guarantee of privacy is naturely ensured by MPC. $\textsf{HashVeri}_{\textsf{MD}}^{\text{mali}}$ ensures strict rigor of AoI due to the Merkle-Damgård construction, where the collision-resistant property of cryptographic hash functions ensures that any falsification of input data $\mathcal{D}$ will lead to a verification failure with overwhelming probability (see Appendix \ref{appendix:collision_resis}). The AoI guarantee of $\textsf{HashVeri}_{\textsf{AP}}^{\text{mali}}$ is based on the MHT structure, which guarantees that falsifying any challenged block will also lead to a verification failure with overwhelming probability. The approximation factor $(\alpha, \beta)$ can be adjusted by tuning the number of challenged blocks $c$, allowing for a trade-off between efficiency and rigor. Our work supports arbitrary utility functions $f()$ defined over $\mathcal{D}$ and $\mathcal{T}$, and works with any MS-MPC protocol that supports binary circuit computation.

\vspace{-0.1in}
\section{Private KNN-Shapley}
\label{sec:private_knn_shapley}

In scenarios involving multiple sellers (contributors), it is critical to equitably reflect each contributor's impact on the overall dataset utility for fair pricing or revenue allocation \cite{agarwal2019marketplace}. To this end, we study the the concrete design and optimization of MPC-based KNN Shapley value calculation to enable fair allocation without disclosing individual data in the non-TCP setting.

\noindent{\textbf{Why KNN-Shapley?}} Despite the variety of Shapley methods and task utilities, we focus on KNN-Shapley for the following reasons:
\begin{itemize}
    \item Enhanced efficiency. The calculation reduces the complexity of standard Shapley from $O(2^N)$ to $O(N \log N)$ (for a single data point) or $O(N^K)$ (for multiple data points), where $N$ denotes the number of participants. Meanwhile, its core operation sorting enables further optimization in the MPC layer, as we will illustrate later.
    \item Comparable accuracy. A KNN classifier can serve as a proxy model for target learning algorithms, simplifying the utility evaluation of the complex models to a leaner KNN model while achieving comparable performance \cite{jia2019efficient,jia2021scalability,wang2025data}. It has been applied across various tasks \cite{ghorbani2022data, shim2021online, liang2020beyond, karlavs2023data,wang2025data}.
\end{itemize}

\vspace{-0.1in}
\subsection{Concrete Design of Private KNN-Shapley}

\subsubsection{KNN Shapley for a Single Data Point}
\label{sec:one_data_per_seller}

We start from a case where the KNN Shapley value is measured for each data points. Based on the KNN-Shapley algorithm that operates on the plain data \cite{jia2019efficient} 
with $O(N\log N)$ complexity, we illustrate the algorithm to calculate the MPC-based KNN-Shapley in Alg. \ref{alg:mpc_exact_knnsv_one}. 

First, each party shares its data (line 1) in secret-shared form to protect privacy in subsequent computation. 
Then, parties obliviously compute the pair-wise distance $\share{d}$ (lines 2-4).
For each test data point $i$, the label $\share{y}$ is permuted based on the vector of permutation index $\share{\pi}$, which represents the secret-shared rank index of $\share{d_i}$ (lines 6-7). Next, all Shapley values can be obtained by a linear scan through all $N$ data points (lines 8-10). The correctness of Alg. \ref{alg:mpc_exact_knnsv_one} follows directly from the plain algorithm \cite{chen2022selling}. The security is naturally clarified because all intermediate data is in secret-shared form and processed by MPC primitives.

\begin{algorithm} \small
\caption{MPC-based algorithm for calculating the KNN Shapley for single data points.}
\label{alg:mpc_exact_knnsv_one}
\KwIn{Selling data $\mathcal{D} = \{(x_j, y_j)\}_{j=1}^N$, test data $\mathcal{T} = \{(x_{\text{test}, j}, y_{\text{test}, j})\}_{j=1}^{N_{\text{test}}}$}
\KwOut{The SV $\{sv_i\}_{i=1}^{N}$}
$\share{\mathcal{D}} \gets \textsf{Share}(\mathcal{D})$, $\share{\mathcal{D}_{\text{test}}} \gets \textsf{Share}(\mathcal{D}_{\text{test}})$ 

\For{$j \gets 1$ \KwTo $N_{\text{test}}$}{ 

    \For{$i \gets 1$ \KwTo $N$}{
        $\share{d_{j,i}} \gets \textsf{ObliDist}(\share{x_i},\share{x_{\text{test},j}})$ 
    }
}

\For{$j \gets 1$ \KwTo $N_{\text{test}}$}{ 

    $\share{\pi} \gets \textsf{ObliSortPerm}(\share{d_j})$
    
    $\share{\hat{y}} \gets \textsf{ApplyPerm}(\share{\pi}, \share{y})$
    
    $\share{sv_{j,N}} \gets \frac{1}{N} \textsf{Eq}(\share{\hat{y}_N}, \share{y_{\text{test},j}})$
    
    \For{$i \gets N-1$ \KwTo $1$}{
        $\share{sv_{j,i}} \gets \share{sv_{j,i+1}} + \frac{1}{K \cdot i} \textsf{Eq}(\share{\hat{y}_i}, \share{y_{\text{test},j}}) \cdot \min(K,i)$
    }
}

\For{$i \gets 1$ \KwTo $N$}{
    $\share{sv_i} \gets \frac{1}{N_{\text{test}}} \sum_{j=1}^{N_{\text{test}}} \share{sv_{j, i}}$
    $sv_i \gets \textsf{Recover}(\share{sv_i})$
}

\Return{$\{sv_i\}_{i=1}^{N}$}
\end{algorithm}

\partitle{Optimization}
A crucial step in Alg. \ref{alg:mpc_exact_knnsv_one} is to obliviously sort $\share{y}$ based on the sorting permutation of $\share{d_j}$. To prevent information leakage, the permutation must be obtained in the secret-shared form via an oblivious sorting permutation protocol $\textsf{ObliSortPerm}$ (illustrated in Alg. \ref{alg:obli_sort_perm}). As suggested in \cite{RadixSort}, this can be instantiated by invoking an oblivious sorting protocol $\textsf{ObliSort}$ with concatenated sorting key $k' = k||\text{index}$ (line 3), where $\text{index}$ is the row number of bit-length $l_{index}$ that is concatenated to the lowest bits of the original sorting key $k$. After sorting based on $k'$ (line 4), the lowest $l_{index}$ bits can be split out (line 5) as the sorted permutation. The complexity of $\textsf{ObliSortPerm}$ is $O(n(l _v + \log n)\log^2 n)$
if the $\textsf{ObliSort}$ is instantiated with wide-adopted Bitonic sort \cite{BitonicSort},
where $n$ is the size of data and $l_v$ is the bit-length of sorting keys. To reduce this overhead, we observe that $\textsf{ObliSortPerm}$ can be instantiated more efficiently with a variant of oblivious Radix sort \cite{RadixSort}. Recall that the oblivious Radix sort iterates each bit and internally generates a secret-shared permutation in each iteration step. It suffices to use the permutation in the final iteration as our desired permutation $\pi$ (line 9). Thus, this approach avoids explicit index concatenation and the complexity of $\textsf{ObliSortPerm}$ can be reduced to $O(nl_v \log n)$. Moreover, $\textsf{ObliDist}$ aims to obliviously compute the L2 distance for ranking. We improve its efficiency by omitting the final square root operation, as it is a monotonic function that does not alter the distance ranking. 

\begin{algorithm}\small
\caption{\textsf{ObliSortPerm}}
\label{alg:obli_sort_perm}
\KwIn{$\share{d}$ of length $n$}
\KwOut{$\share{\pi}$ of length $n$}

// The original version.

\For{$i \gets 1$ \KwTo $n$}{ 
    $\share{d_i'} \gets \textsf{concat}(\share{d_i}, \share{i})$ \hfill  $\triangleright$ $d_i'\gets d_i||j$
}

$\share{d''} \gets \textsf{ObliSort}(\share{d'})$
    
\For{$i \gets 1$ \KwTo $n$}{ 
    $(\share{d'''},\share{\pi_i}) \gets \textsf{Split}(\share{d''}) $ 
}
\Return{$\share{\pi}$}

// The optimized version.

$\share{\pi} \gets \textsf{ObliRadixPerm}(\share{d})$ 
    
\Return{$\share{\pi}$}

\end{algorithm}

\subsubsection{KNN Shapley for General Case}
\label{sec:knn_shapley_general}

In the general case, the KNN Shapley value is measured on a dataset. In this context, the challenge of efficiently computing KNN Shapley under MPC is effectively reduced to optimizing the KNN utility calculation. This is because the Shapley calculation framework only defines the manner in which utility calculations are invoked, without introducing additional privacy risks. 
Consequently, we focus solely on evaluating the efficiency of the KNN utility function.

We present the oblivious KNN Utility algorithm $\textsf{ObliUtil}_{\text{KNN}}$ in the Appendix \ref{appendix:knn_shapley_general} due to space constrains. The core step of it is the oblivious $\textsf{ObliDist}$ and $ \textsf{ObliSortPerm}$. The optimizations proposed in \S\ref{sec:one_data_per_seller} also applies to improve efficiency.

\vspace{-0.1in}
\section{Experiments}
\label{sec:experiments}

In this section, we examine designed protocols in detail. Specifically, we address the following research questions.

\begin{itemize}
    \item \textbf{RQ1:} Do the proposed hash verification protocols effectively provide a trade-off between efficiency and the rigor of guarantee?
    \item \textbf{RQ2:} How about the overhead of the signaling protocol with respect to the security guarantees it provides? Do the proposed signaling protocols efficiently and accurately offer utility value for different tasks and settings? 
    \item \textbf{RQ3:} Is the optimized KNN Shapley protocol efficient compared with the baseline?
\end{itemize}

\subsection{Experiment Setup}

We implement our protocol with the widely-adopted MPC library MP-SPDZ \cite{mp-spdz}. Binary circuit operations (including SHA512 and LowMC) are instantiated using Bristol-style circuit files. As required by the algorithms, signed fixed-point arithmetic is used with a uniform bit length of 64 bits for all values. We simulate a setting with three sellers and one buyer, leveraging the protocol from \cite{dalskov2021fantastic} available in MP-SPDZ to simulate the multi-party environment. Data are split with equal size among sellers. Since the original protocol in \cite{dalskov2021fantastic} only works on MS-MPC model, we remove the malicious checking procedure to adapt it to the SH-MPC setting. All experiments are executed on a machine with Intel$^\circledR$ Core$^{\text{TM}}$ i9-9900K 3.60GHz CPU and 128GB RAM, emulating a LAN network environment with a bandwidth of 2.5Gbps and 0.25ms RTT.

\partitle{Dataset setting} For the efficiency evaluation in \S\ref{sec:eva_hash_veri}, \S\ref{sec:full_efficiency}, and \S\ref{sec:eva_shapley}, we use synthesized datasets with 10 features of various sizes to enable fine-grained observation on efficiency. For the accuracy evaluation in \S\ref{sec:full_accuracy}, we use the Wisconsin Breast Cancer (BCW) dataset \cite{Wolberg1992Breast} for decision tree (DT) and the MNIST dataset \cite{Lecun2010MNIST} for LR and neural network (NN), to evaluate on both low-dimensional tabular and high-dimensional image-based data. The ratio of the test dataset $\mathcal{T}$ is set to 20\%.

\subsection{Hash Verification Protocols}
\label{sec:eva_hash_veri}

To answer RQ1, we evaluate the performance of our hash verification protocols that ensure AoI of the signaling protocol. $\textsf{HashVeri}^{\text{mali}}_{\textsf{MD}}$ (Alg. \ref{alg:hash_veri_md}) and $\textsf{HashVeri}^{\text{mali}}_{\textsf{AP}}$ with $(\alpha,\beta)$-AoI (Alg. \ref{alg:hash_veri_ap}) are simply denoted as \textsf{MD} and \textsf{AP}-$(\alpha,\beta)$, respectively.

\subsubsection{Performace Comparision of \textsf{MD} and \textsf{AP}}

First, we compare the evaluation time of our two protocols with data of various sizes. The underlying compression function $\textsf{H}$ is universally set as LowMC for fair comparison. The time mainly includes MPC-based hash evaluation and MHT operations.

\begin{figure}[t!]
    \centering
    \begin{subfigure}[b]{0.22\textwidth} 
        \centering
        \includegraphics[width=\linewidth]{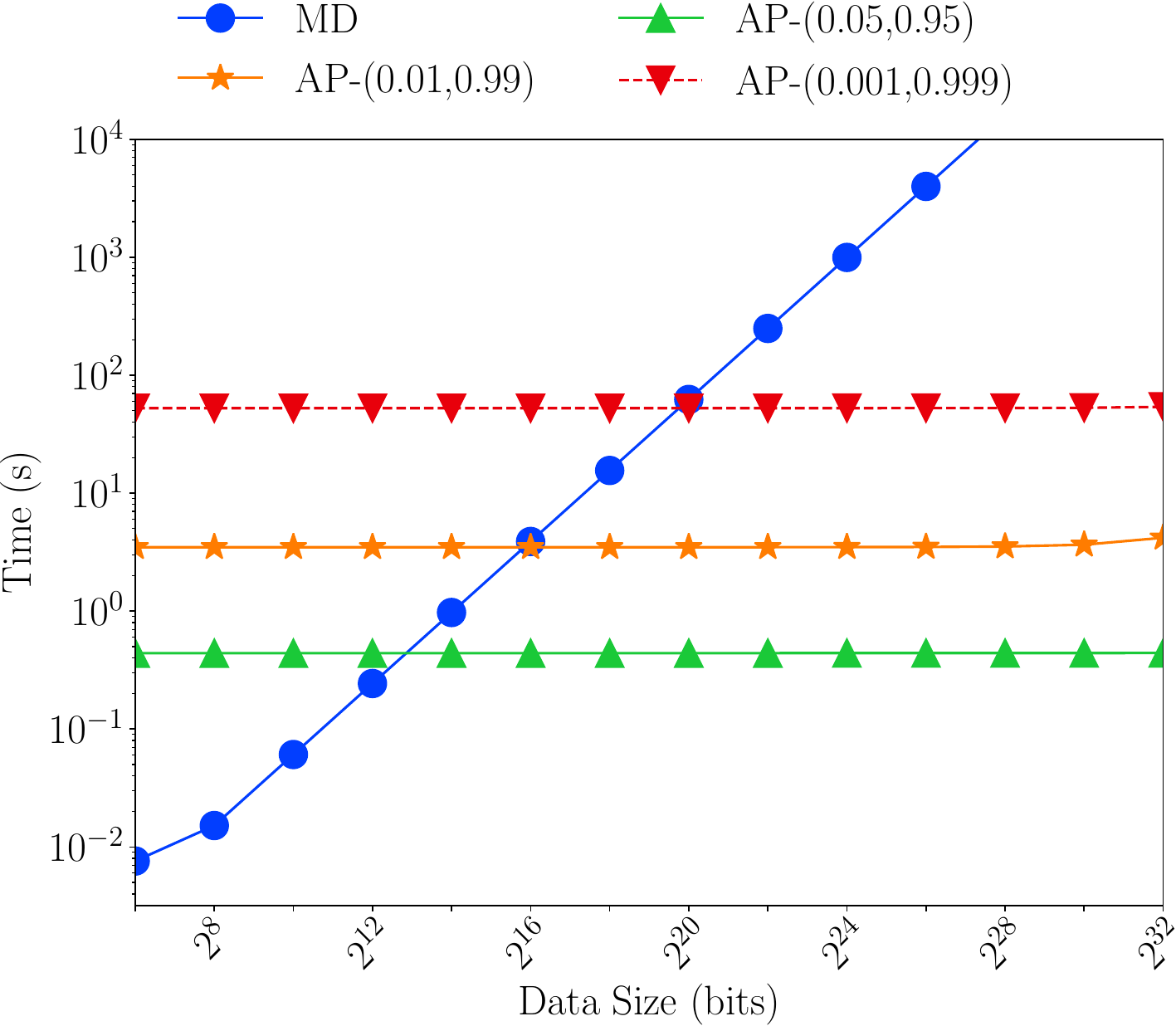} 
        \vspace{-0.26in}
        \caption{Execution time.}
        \label{fig:time_hash}
    \end{subfigure}
    \hspace{0.1in}
    \begin{subfigure}[b]{0.23\textwidth} 
        \centering
        \includegraphics[width=\linewidth]{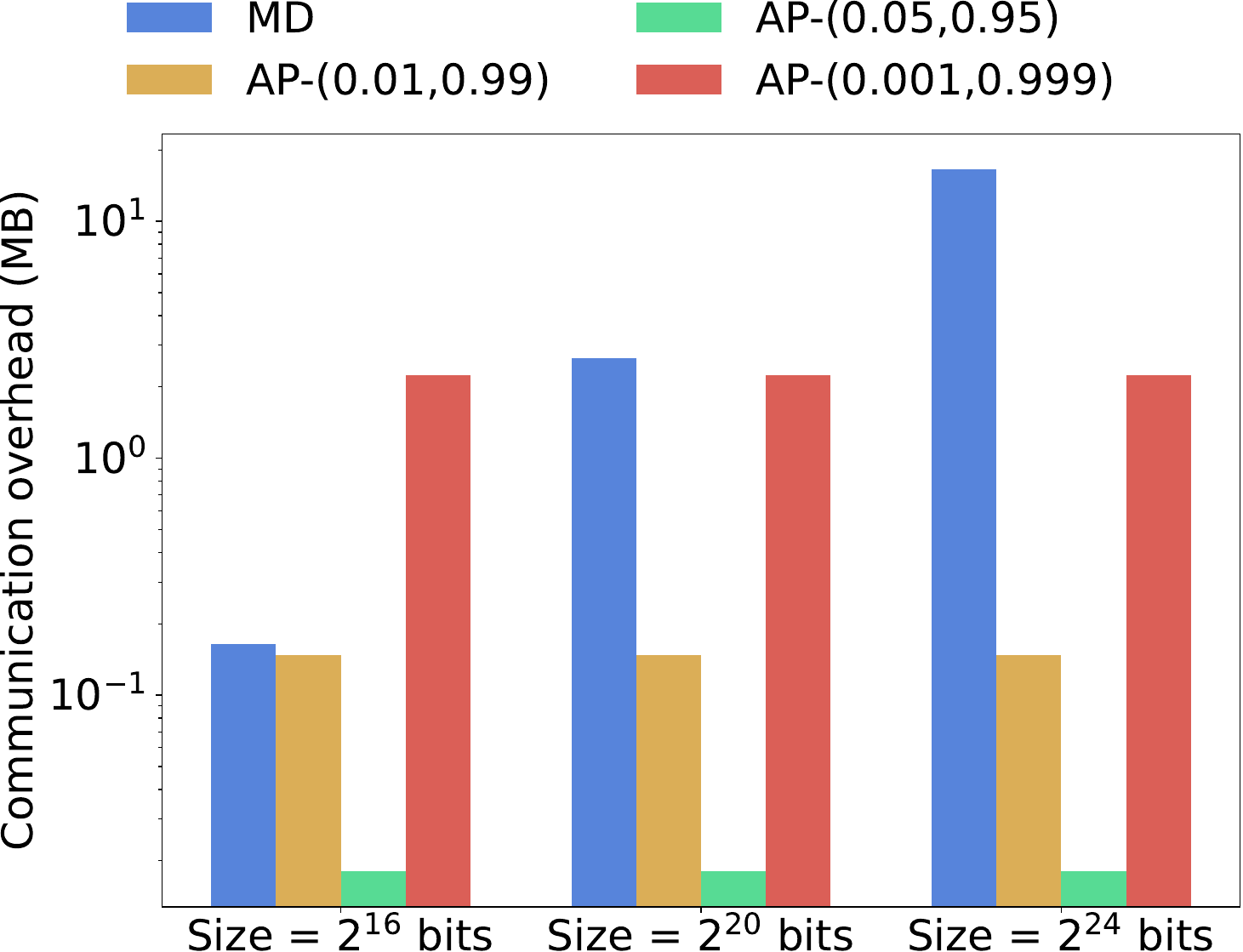} 
        \caption{Communication.}
        \label{fig:commu_hash}
    \end{subfigure}
    \vspace{-0.1in}
    \caption{Execution time and communication overhead of hash verification protocols.}
    \vspace{-0.1in}
    \label{fig:combined_hash}
\end{figure}

Fig. \ref{fig:combined_hash} shows \textsf{MD} that offers a strict guarantee is efficient for smaller datasets (\eg under 1 second for up to $2^{16}$ bits). However, its runtime exhibits linear growth with data size, mainly because the number of $\textsf{AND}$ gates that dominate computation and communication scale linearly (also suggested in Fig. \ref{fig:commu_hash}). Meanwhile, \textsf{AP} significantly reduces overhead by sampling and achieves an almost \textit{constant} execution time. This is due to the number of challenged blocks that require MPC evaluation remaining constant (see \S\ref{sec:hash_ver_pto}), and MHT operations enjoy logarithmic overhead that accounts for a negligible fraction of the total time. In particular, \textsf{AP}-$(0.01,0.99)$ that tolerates at least $1\%$ falsification ensures a $99\%$ guarantee to detect data falsification for $\textit{any}$ size of the data within one minute ($c = 458$), thus allowing scalability to large-scale data.

\subsubsection{Discussion on the Scalability of \textsf{AP}}

\begin{figure}[t!]
    \centering
    \includegraphics[width=0.3\linewidth]{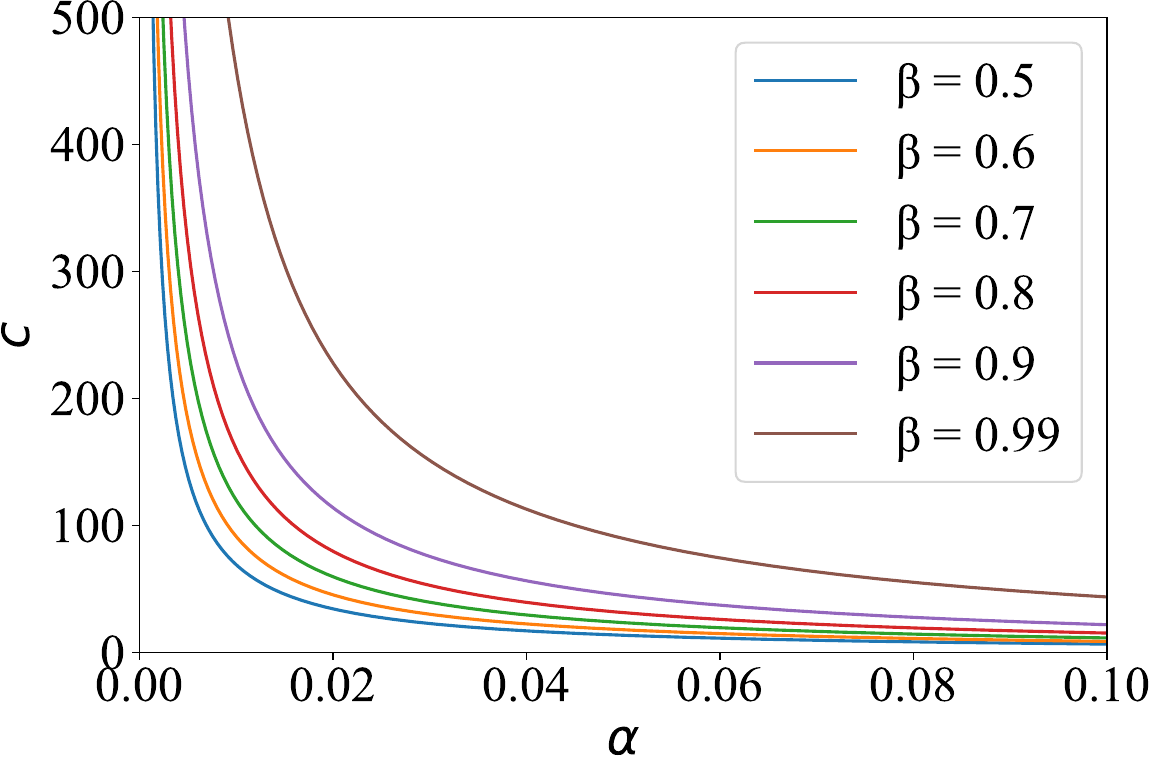} 
    \vspace{-0.1in}
    \caption{The relationship between $c$, $\alpha$ and $\beta$.}
    \vspace{-0.2in}
    \label{fig:approx}
\end{figure}

To further study the scalability of \textsf{AP}, we illustrate the relationship among challenged blocks $c$ that vary with respect $\alpha$ and $\beta$ in Fig. \ref{fig:approx}. Our results show that a relaxed setting of $\alpha$ (\eg > 0.01) and $\beta$ (\eg < 0.8) reduces the number of evaluated blocks to fewer than 200, enabling verification to be completed within seconds. Notably, this relaxed setting may still meet practical requirements, as it is challenging for the seller to manipulate the utility by modifying a small amount of data without knowledge of the distribution of buyer's test data \cite{jagielski2018manipulating}.

\subsubsection{Evaluation of $\textsf{H}$}
We benchmark different instantiations of $\textsf{H}$ of AES-128, SHA-256, SHA-512, and LowMC-128 on $2^{20}$ bits of data, where the suffix denotes the operational block size. The result is shown in Tab. \ref{tab:eva_h}. The MPC-friendly cipher LowMC significantly outperforms standard alternatives, achieving at least $2.29\times$ improvements of throughput over baselines. This improvement is primarily attributed to the reduced communication overhead associated with MPC evaluation of $\textsf{AND}$ gates.

\vspace{-0in}
\begin{table}[ht]

    \centering
    \resizebox{0.45\columnwidth}{!} {
    \begin{tabular}{c|c|c|c|c}
        \toprule
        \textbf{$\textsf{H}$} & AES-128 & SHA-256 & SHA-512 & LowMC-128 \\
        \midrule
        \textbf{Time (s)} & 143.85 & 529.408 & 679.93 & 62.25 \\
        \bottomrule
    \end{tabular}
    }
    \caption{Evaluation of different instantiation of $\textsf{H}$.}
    \vspace{-0.3in}
    \label{tab:eva_h}
\end{table}

\subsection{The Full Signaling Protocol}
\label{sec:exp_signaling}

To answer RQ2, we verify the performance of the full signaling protocol $\Pi_{\mathcal{M}}^{\text{mali}}$.

\partitle{Utility function setting} We instantiate the task-specific utility function $f$ in $\Pi_{\mathcal{M}}^{\text{mali}}$ with three representative tasks drawn from our real-world examples (see §1 and Appendix \ref{appendix:more_signal_examples}). The first, precision marketing (PM), computes the set intersection cardinality: $|\mathcal{D} \cap \mathcal{T}|$. The second, credit risk evaluation (CRE), calculates the number of items in the intersection whose scores exceed a threshold: $\sum_{id \in \mathcal{D}\cap\mathcal{T}} \mathbbm{1}(\textsf{score}_{id} > t)$. Both PM and CRE are implemented using the sort-compare protocol of \cite{huang2012private} due to its compatibility with the arithmetic modules in MP-SPDZ. The third task, model performance validation (MPV), involves performing model inference on the model that is trained on $\mathcal{D}$. The model can be trained locally or by federated learning \cite{mcmahan2017communication}, and we omit the training time for fair comparison. We employ a logistic regression (LR) for this task. 

\partitle{Baseline setting} As this paper presents the first effort to design a reliable and private signaling mechanism, we benchmark our protocol,$\textsf{FullPto}$, against three baselines with varying security guarantees to evaluate the inherent tradeoff between protection and efficiency. Among them, $\textsf{FullPto}$ (Alg. \ref{alg:full_mechanism}) achieves the full protection of privacy and reliability. The first baseline, $\textsf{RoCPto}$, aims to protect privacy and prevent any computation deviation, implemented using an MS-MPC-based utility calculation ($\Pi^{\text{mali}}_f$). The second, $\textsf{AoIPto}$, guarantees privacy and input correctness by combining SH-MPC-based utility calculation ($\Pi^{\text{semi}}_f$) with a MS-MPC-based hash verification ($\textsf{HashVeri}^{\text{mali}}$). The third, $\textsf{PriPto}$, offers minimal protection by only ensuring privacy via the basic protocol in Alg. \ref{alg:signaling_semi} ($\Pi^{\text{semi}}_{\mathcal{M}}$). A summary of these protocols is provided in Tab. \ref{tab:protocol_settings}.

\vspace{-0in}
\begin{table}[ht]
    
    \centering
    \resizebox{0.6\columnwidth}{!} {
        \begin{tabular}{c|c|c|c|c}
        \toprule
\multirow{3}{*}{\textbf{Protocols}} & \multicolumn{3}{c|}{\textbf{Achieved Properties}}                       & \multirow{3}{*}{\textbf{Implementations}}            \\
\cline{2-4}
                          & \multirow{2}{*}{\textbf{Privacy}}  & \multicolumn{2}{c|}{\textbf{Reliability}} &                                             \\
\cline{3-4}
                          &                           & \textbf{RoC}            & \textbf{AoI}            &                                             \\
\midrule
$\textsf{FullPto}$ (Ours)       & \solidCircle              & \solidCircle   & \solidCircle   & $\Pi^{\text{mali}}_{\mathcal{M}}$                \\
$\textsf{RoCPto}$                  & \solidCircle              & \solidCircle   & \openCircle    & $\Pi^{\text{mali}}_f$                     \\
$\textsf{AoIPto}$                  & \solidCircle              & \openCircle    & \solidCircle   & $\Pi^{\text{semi}}_f$ \& $\textsf{HashVeri}^{\text{mali}}$  \\
$\textsf{PriPto}$ (The basic protocol in \S5.1)    & \solidCircle              & \openCircle    & \openCircle    & $\Pi^{\text{semi}}_\mathcal{M}$                           \\              
\bottomrule
\end{tabular}
    }
    \caption{Baseline settings.}
    \vspace{-0.3in}
    \label{tab:protocol_settings}
\end{table}

\subsubsection{Trade-off between Protection Level and Efficiency}
\label{sec:full_efficiency}

We benchmark the efficiency of our signaling protocol against baselines with varying levels of guarantees, showing the results in Tab. \ref{tab:time_full_pto}. For the protocols that include \textsf{HashVeri} (\textsf{FullPto} and \textsf{AoIPto}), we evaluate on both \textsf{MD} and \textsf{AP}-$(0.05,0.95)$. We observe that \textsf{FullPto} exhibits a linear growth in runtime, establishing the performance upper bound as expected. The runtime of \textsf{RoCPto} shows a huge gap between it and $\textsf{FullPto}$, indicating that the cost of the hash verification protocol is the dominant performance bottleneck of \textsf{FullPto-MD}, which is confirmed by up to 99.9\% of $r_{\textsf{MD}}$. In contrast, by utilizing our more efficient \textsf{AP} protocol, the efficiency of \textsf{FullPto-AP} is significantly improved, with $r_{\textsf{AP}}$ being lower than 10\% for most cases, demonstrating the effectiveness of our design. The performance gap between $\textsf{RoCPto}$ and $\textsf{PriPto}$ stems from the cost of malicious checks in underlying MS-MPC modules of \cite{dalskov2021fantastic}, primarily driven by the number of communication rounds. $\textsf{PriPto}$, ensuring only privacy, consistently serves as the fastest baseline.

\noindent\textbf{Remark.} The core design of our solution, hash verification protocols, incurs only a small portion of the total signaling cost (lower than 40\%). Notably, \textsf{AP} reduces the cost to almost a negligible level (as low as 0.01\%), and the improvement is even pronounced as the data size increases. The vast majority of the total cost is incurred by the MPC evaluation of the utility function $f$ of tasks, whose complexity varies with tasks and is anticipated to be improved with ongoing advancements in MPC technologies \cite{islam2025detecting,boyle2025improved,boyle2025preprocessing,bitgc2025}. We also remark that the protocol setting can be configured based on the practical need, \eg if parties are trusted not to misbehave in the signaling stage, \textsf{PriPto} that with lower overhead suffices.

\vspace{-0in}
\begin{table}[ht]

    \centering
    \resizebox{0.6\columnwidth}{!} {
    \begin{tabular}{c|c|c|c|c|c|c|c|c|c}
        \toprule
        \multirow{3}{*}{\textbf{$n$}}  & \multirow{3}{*}{\textbf{Tasks}} & \multicolumn{6}{c|}{\textbf{Time (min)}} & \multirow{3}{*}{\textbf{$r_{\textsf{MD}}$}} & \multirow{3}{*}{\textbf{$r_{\textsf{AP}}$}} \\ 
        \cline{3-8}
         & & \multicolumn{2}{c|}{\textbf{\textsf{FullPto}}} & \multirow{2}{*}{\textbf{\textsf{RoCPto}}} & \multicolumn{2}{c|}{\textbf{\textsf{AoIPto}}} & \multirow{2}{*}{\textbf{\textsf{PriPto}}} & \\
         \cline{3-4} \cline{6-7}
          & & MD & AP & & MD & AP & & \\
        \midrule
        \multirow{3}{*}{$2^8$} & PM & 0.23 & 0.09 & 0.07 & 0.23 & 0.07 & 0.07 & 63.92\% & \textbf{7.42\%} \\
         & CRE & 0.25 & 0.09 & 0.09 & 0.23 & 0.07 & 0.07 & 65.31\% & \textbf{7.8\%} \\
         & MPV & 0.81 & 0.03 & 0.01 & 0.81 & 0.03 & 0.01  & 99.9\% & \textbf{19.9\%} \\
        \midrule
        \multirow{3}{*}{$2^{12}$} & PM & 2.76 & 0.17 & 0.17 & 2.75 & 0.16 & 0.16 & 93.90\% & \textbf{4.17\%} \\
         & CRE & 2.77 & 0.18 & 0.17 & 2.76 & 0.17 & 0.16 & 93.68\% & \textbf{4.02\%} \\
         & MPV  & 12.96 & 0.03 & 0.01 & 12.96 & 0.03 & 0.01 & 99.9\% & \textbf{18.9\%}\\
        \midrule
        \multirow{3}{*}{$2^{16}$} & PM & 43.84 & 2.37 & 2.33 & 43.81 & 2.30 & 2.30 & 94.59\% & \textbf{0.30\%} \\
         & CRE & 44.04 & 2.53 & 2.53 & 43.96 & 2.46 & 2.46 & 94.24\% & \textbf{0.28\%} \\
         & MPV  & 207.55 & 0.06 & 0.03 & 207.54 & 0.05 & 0.01 & 99.9\% & \textbf{10.6\%}\\
        \bottomrule
    \end{tabular}
    }
    \caption{Execution time of different protocols, where $n$ is the number of records in the union dataset. $r_{\textsf{MD}}$ and $r_{\textsf{AP}}$ are the proportion of runtime spent on hash verification in \textsf{FullPto-MD} and \textsf{FullPto-AP}, respectively.} 
    \vspace{-0.3in}
    \label{tab:time_full_pto}
\end{table}

\subsubsection{Different ML Utilities}
\label{sec:full_accuracy}
To further observe the accuracy and efficiency of the utility function $f$ in signaling across different models in the MPV task, we conduct experiments on DT, LR, and NN with various parameters. DT is implemented using the protocol \cite{hamada2023efficient} embedded in MP-SPDZ. NN represents a 3-layer dense NN adopted from \cite{patra2021aby2, wagh2019securenn,mohassel2017secureml}, containing two hidden layers (128 nodes each) and a 10-node output layer, using ReLU activation.

The results are summarized in Tab. \ref{tab:task_settings}. Specifically, for the DT model on the BCW dataset, accuracy peaks at 89.47\% with a moderate tree height ($h=3, 5$), but overfitting at $h=10$. On the more complex MNIST dataset, the NN consistently achieves superior accuracy over LR (peaking at 97.57\% vs. 92.43\%). 
These results demonstrate $\textsf{FullPto}$'s flexibility in supporting diverse ML utility functions of signaling, showing its efficiency and accuracy of offering participants' utility signal. 

\vspace{-0in}
\begin{table}[ht]

    \newcolumntype{C}[1]{>{\centering\arraybackslash}p{#1}}
    \centering
    \resizebox{0.35\columnwidth}{!} {
    \begin{tabular}{c|C{1cm}|C{1cm}|C{1cm}|C{1cm}}
        \toprule
        \textbf{Parameters} & \multicolumn{2}{c|}{\textbf{Accuracy}}  & \multicolumn{2}{c}{ \textbf{Time (s)}}   \\
        \cline{2-3} \cline{4-5}
        \multicolumn{1}{r|}{model} & \multicolumn{2}{c|}{\textbf{DT}} & \multicolumn{2}{c}{\textbf{DT}} \\

        \midrule

        h=3 & \multicolumn{2}{c|}{89.47\%} & \multicolumn{2}{c}{8.61}  \\
        h=5 & \multicolumn{2}{c|}{89.47\%} & \multicolumn{2}{c}{14.35}  \\
        h=10 & \multicolumn{2}{c|}{86.84\%} & \multicolumn{2}{c}{28.70}  \\

        \midrule

        \multicolumn{1}{r|}{model} &  \textbf{LR} & \textbf{NN} & \textbf{LR} & \textbf{NN} \\

        \midrule

        b=32,e=5  & 92.43\% &  97.35\%  & 5.29 & 97.01 \\
        b=64,e=5  & 92.04\% &  97.28\%  & 5.31 & 98.35 \\
        b=128,e=5  & 91.30\% &  96.09\%  & 5.32 & 98.26\\
        b=128,e=10  & 92.02\% &  97.11\%  & 5.32 & 98.31 \\
        b=128,e=15  & 92.10\% &  97.45\%  & 5.32 & 98.19 \\
        b=128,e=20  & 92.27\% &  97.57\%  & 5.28 & 96.42 \\
        
        \bottomrule
    \end{tabular}
    }
    \caption{Accuracy and execution time for different models and parameters, where $h$ is the tree height for DT, $b$ and $e$ are the batch size and the number of epochs for LR and NN.}
    \vspace{-0.3in} 
    \label{tab:task_settings}
\end{table}

\subsection{MPC-based KNN Shapley}
\label{sec:eva_shapley}

To answer RQ3, we investigate the efficiency of our MPC-based KNN Shapley. Considering our work represents the first effort in this topic, we set the unoptimized MPC implementation of KNN Shapley and KNN utility as our baseline.
We simulate three contributors and one buyer as before, and use the BCW data set \cite{Wolberg1992Breast} and randomly sample records of various sizes from it with 100 test data points. The task involves both model training and inference for fair comparison, since KNN is a lazy learner and performs inference using the training data directly.

\begin{figure}[t!]
    \centering
    \begin{subfigure}[b]{0.23\textwidth} 
        \centering
        \includegraphics[width=\linewidth]{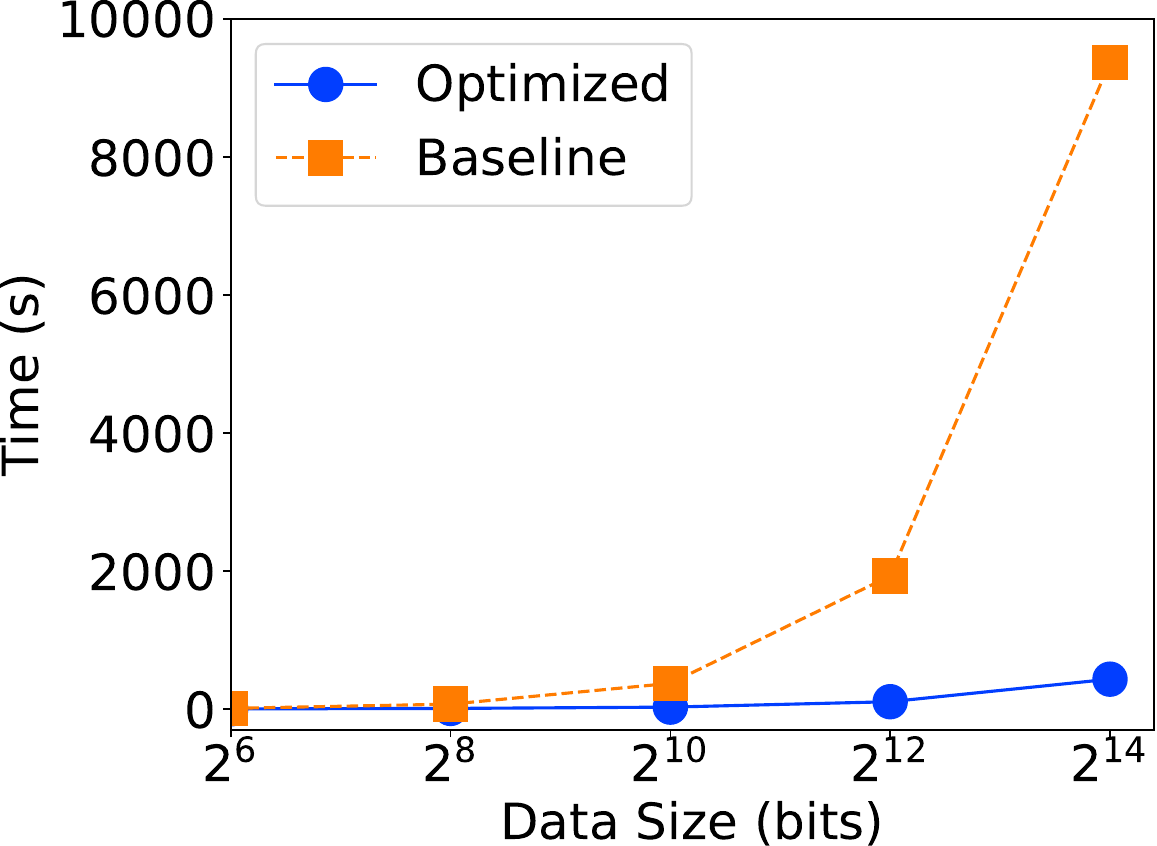} 
        \caption{KNN Shapley for single data points.}
        \label{fig:time_knn_shapley_single}
    \end{subfigure}
    \hspace{0.1in}
    \begin{subfigure}[b]{0.23\textwidth} 
        \centering
        \includegraphics[width=\linewidth]{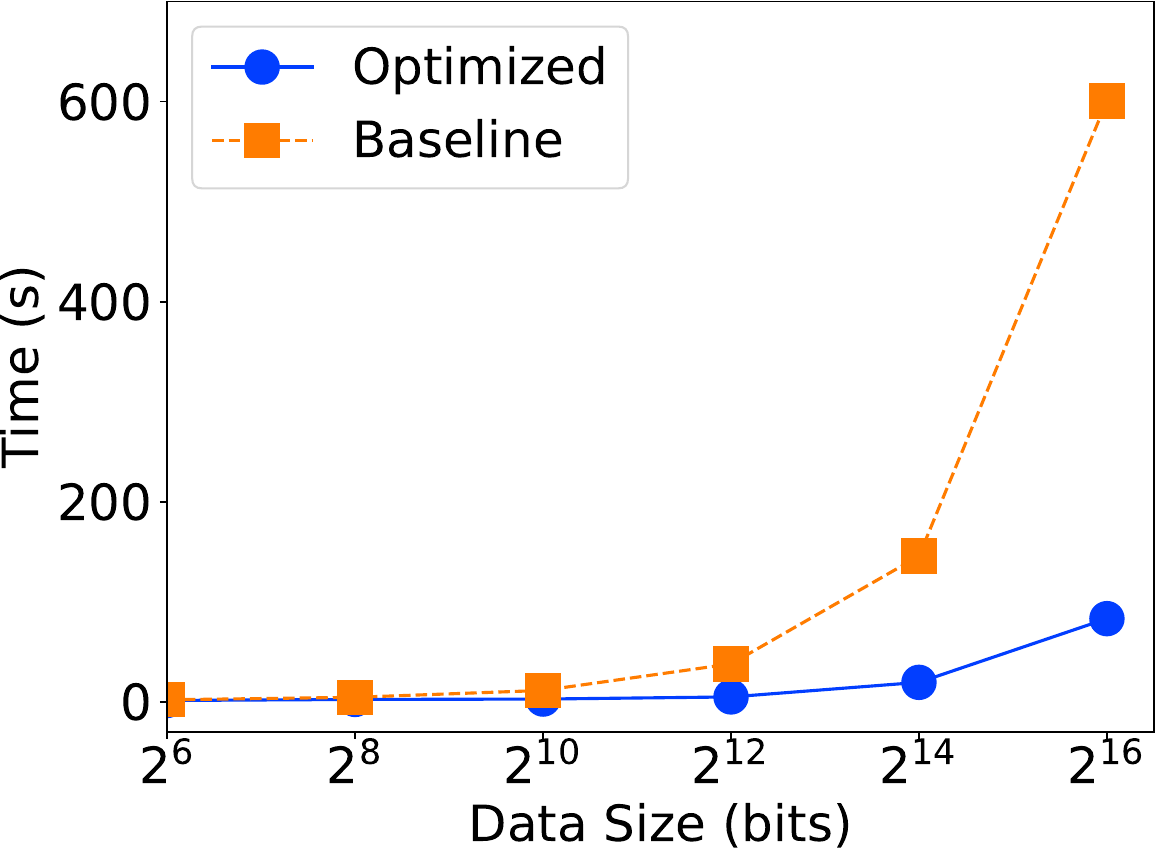} 
        \caption{KNN utility for general Shapley.}
        \label{fig:time_knn_utility}
    \end{subfigure}
    \vspace{-0.1in}
    \caption{Evaluation time for KNN Shapley.}
    \vspace{-0.2in}
\end{figure}

In Fig. \ref{fig:time_knn_shapley_single}, it is demonstrated that our optimized protocol significantly outperforms the baseline. In particular, our optimized approach shows $21.5\times$ faster when the dataset size reaches $2^{14}$ bits and more pronounced with data column growth. The heavy performance of the baseline is mainly attributed to the leveraging of bitonic sort \cite{batcher1968sorting}, which is widely adopted in MPC scenarios but is heavy in evaluation with $O(n(l_v + \log n)\log^2 n)$ complexity. Our exploitation of Radix-based permutation captures the uniqueness of the KNN Shapley setting, obtaining an improved complexity of $O(nl _v\log n)$. Meanwhile, the avoidance of MPC-based square root in \textsf{ObliDist} also improves the efficiency. A similar observation can be obtained in the evaluation of KNN utility for the general case that applies the same optimization, as illustrated in Fig. \ref{fig:time_knn_utility}. 

We further compare the runtime performance of KNN Shapley for the general case against other machine learning tasks, showing in Tab. \ref{tab:eva_shapley}. We use our optimized KNN utility in \S\ref{sec:knn_shapley_general} as the utility function. The results demonstrate that the KNN task outperforms both the DT and NN models, in particular, at $2.2\times$ faster than the NN and nearly $10.2\times$ faster than DT. While the simpler LR model is inherently faster, our optimization makes KNN Shapley more competitive in the MPC setting. This improved efficiency of MPC KNN-Shapley, combined with its high accuracy that is comparable to more complex models \cite{jia2019efficient,jia2021scalability,wang2025data}, broadens its applicability to practical scenarios where no TCP exists.

\vspace{-0in}
\begin{table}[ht]
    \centering
    
    \resizebox{0.27\columnwidth}{!} {
    \begin{tabular}{c|c|c|c|c}
        \toprule
 
        \multirow{2}{*}{\textbf{$n$}} & \multicolumn{4}{c}{\textbf{Time (min)}} \\
        \cline{2-5}
          & \textbf{KNN} & \textbf{DT} & \textbf{LR} & \textbf{NN} \\
        \midrule
        $2^{8}$ & 0.11 & 2.00 & 0.16 & 0.90  \\
        $2^{12}$ & 6.67 & 111.85 & 2.56 & 27.25  \\
        $2^{16}$ & 202.05 & 2079.9 & 42.58 & 458.7 \\
        \bottomrule
    \end{tabular}
    }
    \caption{Efficiency evaluation of Shapley of different ML tasks, where $n$ is the number of data record.}
    \vspace{-0.3in}
    \label{tab:eva_shapley}
\end{table}

\vspace{-0.1in}
\section{Conclusion}
\label{sec:conclusion}

This study highlights the critical role of signaling in data market without relying on TCP. We introduce a signaling mechanism based on MPC, ensuring reliable, private utility signals. This approach benefits both sellers and buyers by facilitating informed decisions. Additionally, we propose an MPC-based KNN-Shapley method for fair utility allocation among multiple sellers. 

\partitle{Future work}  \textbf{(1) Input integrity verification for both parties:} Our work assumes that the buyer prioritizes obtaining the true utility and thus adheres to the protocol. A malicious buyer might input random test data with low utility to mislead the seller into posting a low price. This behavior is challenging to execute and protect against. Execution is difficult because the buyer cannot precisely manipulate the utility by falsifying test data without knowledge of the seller's selling data, potentially causing transaction failure if the resulting price falls below the seller's reservation price, harming both parties' payoff. Protection is also challenging, as assuming a commitment stage for the typically ad-hoc buyer, who lacks a long-term public bulletin, is unrealistic. We leave further exploration of this issue as future work. \textbf{(2) Privacy-preserving Shapley for general tasks:} The Shapley method focused on this work is the KNN Shapley, a method considered to be representative for many general tasks due to its comparable accuracy \cite{jia2019efficient, jia2021scalability, wang2023threshold}. We envision that future methods could bridge this accuracy gap that can compute an \textit{exact} Shapley value in a privacy-preserving way for \textit{any} tasks in the non-TCP setting.

\section*{Acknowledgement}
This work was supported in part by the National Key RD Program of China (2022YFB3103401), NSFC (62102352, 62472378, U23A20306), the Zhejiang Provincial Natural Science Foundation for Distinguished Young Scholars (LR25F020001), and the Zhejiang Province Pioneer Plan (2024C01074, 2025C01084). 

\bibliographystyle{ACM-Reference-Format}
\bibliography{sample}

\clearpage

\appendix

\begin{algorithm}\small
\caption{$\textsf{ObliUtil}_{\text{KNN}}$ }
\label{alg:mpc_knn_util}
\KwIn{Selling data $\mathcal{D} = \{(x_j, y_j)\}_{j=1}^N$, test data $\mathcal{T} = \{(x_{\text{test}, j}, y_{\text{test}, j})\}_{j=1}^{N_{\text{test}}}$}
\KwOut{The secret-shared KNN utility $\share{u}$}

\For{$j \gets 1$ \KwTo $N_{\text{test}}$}{ 

    \For{$i \gets 1$ \KwTo $N$}{
        $\share{d_{j,i}} \gets \textsf{ObliDist}(\share{x_i},\share{x_{\text{test},j}})$ 
    }
}

\For{$j \gets 1$ \KwTo $N_{\text{test}}$}{ 
    $\share{\pi} \gets \textsf{ObliSortPerm}(\share{d_j})$
    
    $\share{\hat{y}} \gets \textsf{ApplyPerm}(\share{\pi}, \share{y})$
    \\
    $\share{u_j} \gets \textsf{Share(0)}$\\
    \For{$i \gets 1$ \KwTo $N$} {
        
        $\share{u_j} \mathrel{+}= \mathsf{Eq}(\share{\hat{y}}, \share{y_{\text{test},j}})$ \hfill  $\triangleright$ equality test \\
        
    }
}
\Return{$\frac{1}{K\cdot N_{\text{test}}}\sum_{j=1}^{N_{\text{test}}}\share{u_j}$}

\end{algorithm}

\section{Frequently Used Notations}
\label{appendix:notation}

We summarize the frequently used notations in Tab. \ref{tab:notation}.

\vspace{-0in}
\begin{table}[ht]

    \centering
    \resizebox{0.7\columnwidth}{!} {
    \begin{tabular}{c|c}
        \toprule
        \textbf{Notations} & \textbf{Description} \\
        \midrule
        $\mathcal{D}$ & The selling dataset hold by the seller\\
        \hline
        $\mathcal{M}$ & The test dataset hold by the buyer \\
        \hline
        $\nu$ & The utility signal \\
        \hline
        $f(\cdot)$ & The utility calculation function \\
        \hline
        $\mathcal{M}$ & The signaling mechanism\\
        \hline
        $\Pi_{\mathcal{M}}$ & The protocol of a signaling mechanism\\
        \hline
        $N$ & Number of records in a dataset \\
        \hline
        $\mathcal{P}_{buyer/seller}$ & The monetary payoff of the buyer/seller \\
        \hline
        $u(\nu,b)$ & The monetary value with respect to the utility $\nu$ and the private type $b$ \\
        \hline
        $x||y$ & The concatenation of two strings $x$ and $y$ \\
        \hline
        $\pi$ & A permutation, which is a bijective function from a finite set to itself \\
         \bottomrule
    \end{tabular}
    }
    \caption{Frequently used notations.}
    \vspace{-0.1in}
    \label{tab:notation}
\end{table}

\section{MAC Checking}
\label{appendix:mac_check}

To check for malicious deviations, we describe a MAC checking mechanism that enhance secret sharing with information-theoretic MACs, as used in SPDZ-like protocols \cite{damgaard2013practical}. An \textit{authenticated share} of a value $x \in \mathbb{F}_p$ held by party $P_i$ is a tuple $\llbracket x \rrbracket_i := ([x]_i, [\alpha \cdot x]_i)$, where $\alpha \in \mathbb{F}_p$ is a secret MAC key, itself also additively shared among the parties. The MAC shares are used in a CheckMAC procedure to verify that the parties correctly computed and revealed a set of values. Instead of checking each value individually, multiple checks are batched into a single one to enhance efficiency using a random linear combination. This procedure ensures that any malicious modification to the opened values is detected with overwhelming probability, without revealing the MAC key $\alpha$.

\noindent \textbf{Protocol CheckMAC} \cite{damgaard2013practical} (for opened values $v_1, \ldots, v_t$):
This protocol is executed after a set of $t$ values have been revealed, resulting in public values $v_1, \ldots, v_t \in \mathbb{F}_p$. For each corresponding secret $\llbracket x_k \rrbracket$, the parties still hold the secret-shared MAC $[\alpha \cdot x_k]$. The goal is to check if $v_k = x_k$ for all $k=1, \ldots, t$.

\begin{enumerate}
    \item Random challenge generation: The parties collaboratively generate a secret-shared random vector $\share{\vec{r}} = (\share{r_1}, \ldots, \share{r_t})$, where each $r_k \in \mathbb{F}_p$ is uniformly random. They then reveal $\vec{r}$ to all participants.
    
    \item Compute public linear combination: All parties locally compute the public value $v_{comb} \in \mathbb{F}_p$, which is the random linear combination of the opened values:
    $$
    v_{comb} := \sum_{k=1}^{t} r_k \cdot v_k
    $$
    
    \item Compute secret-shared MAC combination: Each party $P_i$ locally computes its share of the MAC for the combined secret value. Let $[\gamma(x_k)]_i$ denote the $i$-th share of the MAC for $x_k$ (i.e., $[\alpha \cdot x_k]_i$).
    $$
    [\gamma_{comb}]_i := \sum_{k=1}^{t} r_k \cdot [\gamma(x_k)]_i
    $$
    Due to the linearity of the secret sharing, the parties now hold an additive sharing of $\gamma_{comb} = \alpha \cdot (\sum_{k=1}^t r_k x_k)$.

    \item Compute and open the check value: Each party $P_i$ computes its share of the final check value, $\sigma$:
    $$
    [\sigma]_i := [\gamma_{comb}]_i - v_{comb} \cdot [\alpha]_i
    $$
    The parties then reveal their shares $[\sigma]_i$ and reconstruct the public value $\sigma = \sum_i [\sigma]_i$.

    \item Verification: The protocol succeeds if and only if $\sigma = 0$. If $\sigma \neq 0$, the parties abort.
\end{enumerate}

The security guarantee of \cite{damgaard2013practical} ensures that if at least one value is incorrect (say $v_j \neq x_j$), then the term $\sum r_k(x_k-v_k)$ will be a non-zero polynomial in the random variables $r_k$, which leads to check failure.

\section{Formal Definition of Collision-resistant}
\label{appendix:collision_resis}

The collision resistance is a fundamental concept in cryptography \cite{katz2007introduction}. Its formal definition can be stated as follow:

\begin{definition}
    (Collision-Resistance) Let $ \textsf{H}: \{0,1\}^* \to \{0,1\}^n $ be a function that maps an input of arbitrary length to a fixed-length output of $n$-bits. The function $\textsf{H}$ is said to be collision-resistant if for any probabilistic polynomial-time (PPT) adversary $ \mathcal{A} $, the probability of finding a collision is negligible. Formally:

    $$
    \Pr \left[ x_1 \neq x_2 \land H(x_1) = H(x_2) \;|\; (x_1, x_2) \gets \mathcal{A}(1^\lambda) \right] \leq \text{negl}(\lambda),
    $$
    
    where $ \lambda $ is the security parameter. $ \text{negl}(\lambda) $ denotes a negligible function, \ie a function that decreases faster than the inverse of any polynomial. $ \mathcal{A}(1^\lambda) $ denotes the adversary $ \mathcal{A} $ running in probabilistic polynomial time with access to the security parameter $ 1^\lambda $.
\end{definition}

\section{Proofs of Theorems and Lemmas}

\subsection{Proof of Lemma \ref{lem:unique_solution}}
\label{appendix:proof_lemma1}

\begin{proof}
    The first order condition of problem \eqref{equ:seller_basic_revenue_2} is 
    \begin{equation}
        1 - F_0(p) - pf_0(p) = 0.
    \end{equation}
    So if $p - \frac{1 - F_0(p)}{f_0(p)}$ is monotonically increasing, problem \eqref{equ:seller_basic_revenue_2} has a unique solution, which equals to
    \begin{equation}
    \label{equ:unique_solution_condition}
        \frac{\text{d}\left(p - \frac{1 - F_0(p)}{f_0(p)}\right)}{\text{d} p} = \frac{2f^2_0(p) + f'_0(p)(1 - F_0(p))}{f^2_0(p)} \geq 0.
    \end{equation}
    If $1 - F_0$ is log-concave, which means $\log(1 - F_0)$ is concave, which implies that
    \begin{equation}
    \label{equ:log-concave}
        \frac{\text{d}^2 \log (1 - F_0(p))}{\text{d}p^2} = -\frac{f^2_0(p) + f'_0(p)(1 - F_0(p))}{(1 - F_0(p))^2} \leq 0.
    \end{equation}
    It's obvious that \eqref{equ:log-concave} implies \eqref{equ:unique_solution_condition}, so the seller's payoff maximization problem \eqref{equ:seller_basic_revenue_2} has a unique solution if $1 - F_0$ is log-concave.
\end{proof}

\subsection{Proof of Theorem \ref{thm:seller_rev_inc}}
\label{appendix:proof_theorem1}

\begin{proof}
    If $u(b) > u_0(b)$, then the probability that $u(b) < p$ is smaller than the probability that $u_0(b) < p$, \ie $\forall p, F(p) < F_0(p)$, so $\mathcal{P}_{seller,\mathcal{DT}_{\mathcal{M}}}(p_0^*) > \mathcal{P}_{seller,\mathcal{DT}_{-}}(p_0^*)$. Furthermore, by the optimality definition of $p^*$, we can derive that $\mathcal{P}_{seller,\mathcal{DT}_{\mathcal{M}}}(p^*) \geq \mathcal{P}_{seller,\mathcal{DT}_{\mathcal{M}}}(p_0^*) > \mathcal{P}_{seller,\mathcal{DT}_{-}}(p_0^*)$.
\end{proof}

\subsection{Proof of Lemma \ref{lem:price_decrease}}
\label{appendix:proof_lemma2}

\begin{proof}
    Under the log-concave property of $1 - F_0$ and $1 - F$, the pricing solutions $p_0^*$ and $p^*$ emerge as unique payoff maximizers for optimization problems \eqref{equ:seller_basic_revenue_2} and \eqref{equ:seller_pis_revenue_2} respectively, thus satisfying first order conditions
    \[p_0^* = \frac{1 - F_0(p_0^*)}{f_0(p_0^*)}, \quad p^* = \frac{1 - F(p^*)}{f(p^*)}.\]
    Given $F$ is hazard rate dominated by $F_0$, $\frac{1 - F(p_0^*)}{f(p_0^*)} < \frac{1 - F_0(p_0^*)}{f_0(p_0^*)} = p_0^*$, \ie $p_0^* - \frac{1 - F(p_0^*)}{f(p_0^*)} > 0$. The log-concavity of $1 - F$ ensures the mapping $\phi(p) := p - \frac{1 - F(p)}{f(p)}$ maintains strict monotonicity. Combining this with $\phi(p^*) = 0$, we establish that $p^* < p_0^*$.
\end{proof}

\subsection{Proof of Theorem \ref{thm:buyer_uti_inc}}
\label{appendix:proof_theorem2}

\begin{proof}
    Under the assumption of Lemma \ref{lem:price_decrease}, the relationship $p^* < p^*_0$ holds. We systematically examine four mutually exclusive and collectively exhaustive scenarios:
    \begin{enumerate}
        \item $u_0(b) \geq p_0^*, u(b) \geq p^*$: In this case, both mechanisms induce data purchase, but $\mathcal{DT_{\mathcal{M}}}$ achieves strictly higher buyer utility through price reduction, \ie $\mathcal{P}_{buyer,\mathcal{DT}_{\mathcal{M}}}(p^*) > \\ \mathcal{P}_{buyer,\mathcal{DT}_{-}}(p^*_0)$;
        \item $u_0(b) < p_0^*, u(b) < p^*$: In this case, transaction failure occurs under both mechanisms, yielding identical zero utilities;
        \item $u_0(b) \geq p_0^*, u(b) < p^*$: In this case, because $u(b) < p^* < p^*_0$, $\mathcal{DT_{\mathcal{M}}}$ prevents loss-making transactions that would occur in the basic case $\mathcal{DT_{-}}$, thereby improving utility: $\mathcal{P}_{buyer,\mathcal{DT}_{\mathcal{M}}}(p^*) > \mathcal{P}_{buyer,\mathcal{DT}_{-}}(p^*_0)$;
        \item $u_0(b) < p_0^*, u(b) \geq p^*$: In this case, the buyer would not buy data in basic case $\mathcal{DT_{-}}$ with utility $0$, but would buy data in $\mathcal{DT_{\mathcal{M}}}$ with non-negative utility, so we have $\mathcal{P}_{buyer,\mathcal{DT}_{\mathcal{M}}}(p^*) \geq \mathcal{P}_{buyer,\mathcal{DT}_{-}}(p^*_0)$.
    \end{enumerate}
    The above four cases are collectively exhaustive, given that case 1 occurs with positive measure, we can derive that $\mathcal{P}_{buyer,\mathcal{DT}_{\mathcal{M}}}(p^*) > \mathcal{P}_{buyer,\mathcal{DT}_{-}}(p^*_0)$.
\end{proof}

\subsection{Proof of Theorem \ref{thm:signaling_after_pricing}}
\label{appendix:proof_theorem3}

\begin{proof}
    Note that the price in $\mathcal{DT}_{\mathcal{M}}'$ would remain the same as $\mathcal{DT}_{-}'$ since the seller's information remains unchanged. Given $p$, we examine two mutually exclusive and collectively exhaustive scenarios:
    \begin{enumerate}
        \item When $u_0(b) \geq p$ (\ie $\mathcal{P}_{buyer,\mathcal{DT}_{-}'} \geq 0$), the buyer decides to purchase in the basic case. Since $u(\nu,b)$ is non-decreasing and $p$ is positive, there exists $\nu^*$ such that $u(\nu^*,b) < p $, which leads to $\mathcal{P}_{buyer,\mathcal{DT}_{-}'}^{\text{true}} (\nu^*) < 0$. Since the buyer with a signal $\nu^*$ will refuse to purchase and her true payoff will be zero,  it holds that $\mathcal{P}_{buyer,\mathcal{DT}_{\mathcal{M}}'}^{\text{true}} (\nu^*)> \mathcal{P}_{buyer,\mathcal{DT}_{-}'}^{\text{true}} (\nu^*) $. For other cases where $u(\nu,b) \ge p $, the buyer's decision remains the same, so is their true payoff.
        \item When $u_0(b) < p$ (\ie $\mathcal{P}_{buyer,\mathcal{DT}_{-}'} < 0$),  the buyer refuses to purchase in the basic case. Thus,  $\mathcal{P}_{buyer,\mathcal{DT}_{-}'}^{\text{true}} = 0$. Similar as situation (1), there exists $\nu^*$ such that $u(\nu^*,b) \ge p $, which leads to $\mathcal{P}_{buyer,\mathcal{DT}_{\mathcal{M}}'}^{\text{true}} (\nu^*) > \mathcal{P}_{buyer,\mathcal{DT}'_{-}}^{\text{true}}(\nu^*) = 0$.  For other cases where $u(\nu,b) < p $, the buyer's decision remains the same, so is their true payoff.
    \end{enumerate}
    The above two cases are collectively exhaustive, we can derive that $\mathcal{P}_{buyer,\mathcal{DT}'_{\mathcal{M}}}^{\text{true}} > \mathcal{P}_{buyer,\mathcal{DT}'_{\mathcal{-}}}^{\text{true}}$.
\end{proof}

\subsection{Proof of Theorem \ref{thm:app_hash_veri}}
\label{appendix:proof_theorem4}

\begin{proof}
    We compute $P(X)$, the probability that at least one of the blocks selected by $I$ corresponds to a block falsified by the seller.
    
    \begin{align*}
    P_X &= P\{X \geq 1\} = 1 - P\{X = 0\} \\
    &= 1 - \frac{k-t}{k} \cdot \frac{k-1-t}{k-1} \cdot \frac{k-2-t}{k-2} \cdots \frac{k-c+1-t}{k-c+1}.
    \end{align*}
    
    Since $\frac{k-i-t}{k-i} \geq \frac{k-i-1-t}{k-i-1}$,  it follows that:
    
    \begin{align*}
    1 - \left( \frac{k-t}{k} \right)^c &\leq P_X \leq 1 - \left( \frac{k-c+1-t}{k-c+1} \right)^c.
    \end{align*}
    
    This implies that if the seller falsifies $t$ blocks of data, the buyer has a probability of at least $1 - \left( \frac{k-t}{k} \right)^c$ to detect this misbehavior after challenging $c$ blocks.
\end{proof}

\section{Disscussion on the case where $b$ is Public}
\label{appendix:public_b}

Consider the scenario in which the buyer publicly reveals her type $b$ to the seller. We denote the case by $\mathcal{DT}_{\mathcal{M}}^b$. Given that the data utility $\nu$ is continua, an intuitive example is the buyer publishing her purchasing strategy specified by $b$ that the buyer would accept $p_1$ if $\nu > t_1$, accept $p_2$ if $t_1 > \nu > t_2$ (where $t_1, t_2$ are threshold values) and refuse to purchase otherwise. $\mathcal{DT}_{\mathcal{M}}^b$ is prevalent in practical buyer's market, where the buyer, as the dominant party in the transaction, often proposes the terms of the deal first. 

With $b$ known by the seller, the seller can set the optimal price $p^*=p_{\nu}$ to ensure that the buyer accepts the price while maximizing the seller's payoff. The only remaining task is to signal $\nu$ using $\mathcal{M}$, after which the pricing and purchasing decisions proceed as planned. It is easy to see that $\mathcal{DT}_{\mathcal{M}}^b$ ensures no participant suffers a payoff loss, as the strategy of both parties is already agreed by themselves.

\section{KNN Utility Calculation based on KNN Shapley}
\label{appendix:knn_shapley_general}

We present the algorithm for computing the KNN utility in MPC in Alg. \ref{alg:mpc_knn_util}. The algorithm computes the KNN utility by first calculating the distances between each test data point and all training data points (lines 1-3), then sorting these distances to find the K nearest neighbors (lines 5-6), and finally checking how many of these neighbors match the label of the test data point (lines 8-9).

\section{More Signaling Examples}
\label{appendix:more_signal_examples}

We present two more representative signaling examples with different utility functions as below.

\begin{example}
    (Credit Risk Evaluation) Consider a scenario in which a credit agency (CA), acting as a data seller, maintains a large dataset $\mathcal{D} := (\textsf{idA}, \textsf{score})$ of users’ credit scores derived from their repayment records and risk profiles. A bank (BA), as a data buyer, seeks to obtain the credit score of every loan applicant in its dataset $\mathcal{T} := (\textsf{idB})$ and only permit a loan to high-quality loan applicants (credit score above a threshold $t$). To evaluate the potential value of CA’s data before making a purchase decision, the BA requests a signal - a metric $\nu$ representing the number of its applicants with $\textsf{score} > t$ within the CA’s dataset, \ie $\nu \gets \sum_{id \in \mathcal{D}\cap\mathcal{T}} \mathbbm{1}(\textsf{score}_{id} > t) $. A higher $\nu$ indicates more high-quality loan applicants can be identified by purchasing CA's data and leads to higher business value.
\end{example}

\begin{example}
    (Model Performance Validation) Consider a scenario in which a data provider (DP), acting as a data seller, maintains a labeled dataset $\mathcal{D} := (\mathbf{x}, y)$, collected from a specific domain or application. A machine learning practitioner (MP), serving as a data buyer, aims to evaluate the utility of DP’s data for training a model on their own task. To assess the potential value of DP’s dataset before purchase, MP requests a signal — a performance metric $\nu$, defined as the prediction accuracy (or RMSE, AUC, etc.) of a model $\mathcal{L}$ trained on $\mathcal{D}$ evaluated on MP’s local test set $\mathcal{T} := (\mathbf{x'}, y')$. That is,
    $$
    \nu \gets \text{Accuracy}(\mathcal{L}_{\mathcal{D}}(\mathbf{x'}), y'), \quad \text{for } (\mathbf{x'}, y') \in \mathcal{T}.
    $$
    A higher $\nu$ indicates that DP’s data is more compatible with MP’s task and can lead to better model performance. 
\end{example}

\section{Concrete Example of Buyer's Payoff Improvement} 
\label{appendex:payoff_example}

Let $ u(\nu, b) = \nu \cdot b $ \cite{chen2022selling}. Let $ \nu \sim \text{Uniform}(2, 6) $ and $ b = 2 $ for the buyer. When the price is fixed as $ p = 6 $, the buyer initially expects utility $ u_0(b) = \overline{\nu} \cdot b = 8 > p $, leading to a purchase decision with expected payoff $ \mathcal{P}_{buyer, \mathcal{DT}'_{-}} = 2 $. However, the true utility might be $ \nu^* = 2 $, resulting in a true payoff of $ \mathcal{P}_{buyer,\mathcal{DT}'_{-}}^{true} = 2 \cdot 2 - 6 = -2 $ if purchases. By applying the signaling method $ \mathcal{M} $, which reveals the $\nu^*$ to the buyer before purchasing, the buyer rationally decides not to proceed. Consequently, her true payoff is improved to $\mathcal{P}_{buyer,\mathcal{DT}'_{\mathcal{M}}}^{true} = 0 $.

\end{document}

%% file: header.tex
\usepackage{xcolor}
\usepackage{tabularx}
\usepackage{graphicx}
\usepackage{textcomp}
\usepackage{listings}
\usepackage{array}
\usepackage{flushend}
\usepackage{framed}
\usepackage{amsmath,amsfonts,tikz,multirow,stmaryrd,xspace,enumitem}
\usepackage{url}
\usepackage{amsthm}
\usepackage{caption}
\usepackage{pgfplots}
\pgfplotsset{compat=newest}
\usepackage{ellipsis, ragged2e, marginnote}
\usepackage[]{hyperref}
\hypersetup{hidelinks}
\usepackage{threeparttable}
\usepackage{marvosym}
\usepackage{flushend}
\usepackage[linesnumbered,ruled,vlined]{algorithm2e}
\usepackage{bbm}
\usepackage{algpseudocode}
\usepackage{booktabs}
\usepackage{tikz}
\usepackage{multirow}
\usepackage{subcaption}
\usepackage{makecell}
\newtheorem{theorem}{Theorem}
\newtheorem{remark}{Remark}

\definecolor{mygray}{gray}{.9}
\definecolor{mypink}{rgb}{.99,.91,.95}
\definecolor{mycyan}{cmyk}{.3,0,0,0}
\definecolor{mygreen}{RGB}{1,145,1}
\definecolor{lightgreen}{RGB}{144,238,144}

\def\hashverisha256{\textsf{HashVeri}_{\textsf{SHA256}}}

\def\etal{et al. }
\def\eg{\textit{e.g.}, }
\def\ie{\textit{i.e.}, }

\def\field{\mathbb{F} }
\def\ring{\mathbb{Z}}

\newtheorem{definition}{Definition}

\newtheorem{example}{Example}[]

\newcommand{\share}[1]{\ensuremath{\llbracket #1\rrbracket}\xspace}

\newcommand{\partitle}[1]{\medskip \noindent \textbf{#1.}}

\newcommand{\openCircle}{\tikz\draw (0,0) circle (0.1);}
\newcommand{\solidCircle}{\tikz\fill (0,0) circle (0.1);}

\newtheoremstyle{scit}
  {3pt}                
  {3pt}                
  {\itshape}           
  {}                   
  {\scshape}           
  {.}                  
  {0.5em}              
  {\thmname{#1}~\thmnumber{#2}} 
